\documentclass[12pt]{article}
\usepackage[utf8]{inputenc}
\usepackage{amsmath,amssymb,amsthm}
\usepackage{graphicx}
\usepackage{hyperref}
\usepackage{comment}
\DeclareMathOperator{\E}{\mathbb{E}}

\newcommand{\gM}{\mathcal{M}}

\newcommand{\gG}{\mathcal{G}}
\newcommand{\gB}{\mathcal{B}}
\newcommand{\gS}{\mathcal{S}}

\newcommand{\otheta}{\overline{\theta}}
\newcommand{\uomega}{\underline{\omega}}

\newcommand{\op}{\overline{p}}
\newcommand{\Beta}{B}
\newcommand{\hgM}{\widehat{\gM}}
\newcommand{\hq}{\widehat{q}}
\newcommand{\htr}{\widehat{t}}
\newcommand{\hbeta}{\widehat{\beta}}
\newcommand{\hsigma}{\widehat{\sigma}}

\newcommand{\hgB}{\widehat{\gB}}
\newcommand{\hgS}{\widehat{\gS}}
\newcommand{\eqsymbol}{\rho}
\newcommand{\etsymbol}{\tau}
\newcommand{\eusymbol}{\nu}
\newcommand{\evsymbol}{\nu}
\newcommand{\eqBo}{\eqsymbol^{B0}}
\newcommand{\eqBi}{\eqsymbol^{B1}}
\newcommand{\eqS}{\eqsymbol^S}
\newcommand{\etBo}{\etsymbol^{B0}}
\newcommand{\etBi}{\etsymbol^{B1}}
\newcommand{\etS}{\etsymbol^S}
\newcommand{\euBo}{\eusymbol^{B0}}
\newcommand{\euBi}{\eusymbol^{B1}}
\newcommand{\evS}{\evsymbol^S}
\newcommand{\heqBo}{\widehat{\eqsymbol}^{B0}}
\newcommand{\heqBi}{\widehat{\eqsymbol}^{B1}}
\newcommand{\heqS}{\widehat{\eqsymbol}^S}

\newcommand{\hetBi}{\widehat{\etsymbol}^{B1}}

\newcommand{\heuBo}{\widehat{\eusymbol}^{B0}}
\newcommand{\heuBi}{\widehat{\eusymbol}^{B1}}
\newcommand{\hevS}{\widehat{\evsymbol}^S}
\newcommand{\hBeta}{\widehat{\Beta}}
\newcommand{\hSigma}{\widehat{\Sigma}}
\newcommand{\one}{\mathbf{1}}
\newcommand{\up}{\underline{p}}
\newcommand{\oq}{\overline{q}}
\newcommand{\ot}{\overline{t}}

\newcommand{\hP}{\widehat{P}}
\newcommand{\hV}{\widehat{V}}
\newcommand{\thetaone}{\widetilde{\theta}}
\newcommand{\deltastatic}{\overline{\delta}}

\newtheorem{taggedlemmax}{Lemma}
\newenvironment{taggedlemma}[1]{\renewcommand\thetaggedlemmax{#1}\taggedlemmax}{\endtaggedlemmax}
\title{Pricing Novel Goods\thanks{
We are grateful to
Rossella Argenziano,
Laura Doval,
Duarte Gonçalves,
Johannes Hörner,
Andreas Kleiner,
Alessandro Pavan,
Ludvig Sinander,
Vasiliki Skreta,
and
seminar participants at the
University of Bristol, 
University of Nottingham,
University of Essex, 
Queens University Belfast, 
University of Copenhagen,
University of Manchester,
University of Leicester,
University of Bath,
Cardiff University,
Durham Economic Theory Conference, 
Birmingham Economic Theory Workshop, 
SAET (Paris), 
Junior Workshop in Economic Theory (Bonn), 
Lancaster Game Theory Conference, 
QMUL/City Workshop in Economic Theory (London),
CEPS PolEcon Workshop (Paris),
Conference on Mechanism and Institution Design (Budapest) 
for their comments and suggestions.
A one-page abstract of an earlier version of this paper was published in Proceedings of the 24th ACM Conference on Economics and Computation (EC ’23) \url{https://doi.org/10.1145/3580507.3597694}.
}%
}
\author{Francesco Giovannoni%
\thanks{University of Bristol, \href{mailto:francesco.giovannoni@bristol.ac.uk}{\url{francesco.giovannoni@bristol.ac.uk}}}%
\and
Toomas Hinnosaar%
\thanks{University of Nottingham and CEPR, \href{mailto:toomas@hinnosaar.net}{\url{toomas@hinnosaar.net}}}%
}
\usepackage{datetime}
\newdateformat{monyear}{\monthname[\THEMONTH] \THEYEAR} 
\date{\monyear\today%
}
\usepackage{float}
\usepackage{anysize}
\usepackage{setspace}
\usepackage[titletoc,title]{appendix}

\newtheorem{corollary}{Corollary}

\newtheorem{definition}{Definition}
\newtheorem{lemma}{Lemma}
\newtheorem{proposition}{Proposition}

\usepackage[justification=centering,textfont={sc},labelfont={sc}]{caption,subcaption}
\usepackage{apptools}
\AtAppendix{
  \counterwithin{lemma}{section}
  \counterwithin{remark}{section}
  \counterwithin{corollary}{section}
}
\usepackage{hyperref}
\usepackage{cleveref}
\crefname{figure}{figure}{figures}
\crefname{equation}{equation}{equations}
\usepackage{enumerate}
\usepackage[round, longnamesfirst]{natbib}
\newcommand{\expect}{\mathbb{E}}
\newcommand{\expectG}{\expect_G}

\DeclareMathOperator{\supp}{supp}
\DeclareMathOperator{\argmax}{argmax}
\newcommand{\R}{\mathbb{R}}

\usepackage{todonotes}

\begin{document}

\maketitle
\begin{abstract}
We study a bilateral trade problem where a principal has private information that is revealed with delay, such as a seller who does not yet know her production cost. Postponing the contracting process incurs a costly delay, while early contracting with limited information can create incentive issues, as the principal might misrepresent private information that will be revealed later. We show that the optimal mechanism can effectively address these challenges by leveraging the sequential nature of the problem. The optimal mechanism is a menu of two-part tariffs, where the variable part is determined by the principal's incentives and the fixed part by the agent's incentives. As two-part tariffs might be impractical in some applications, we also study price mechanisms. We show that the optimal price mechanism often entails trade at both the ex-ante and ex-post stages. Dynamic price mechanisms can lower the cost of delay by transacting with high-type agents early and relax the incentive constraints by postponing contracts with lower-type agents. We also generalize our analysis to costly learning and study ex-post efficiency in our context.
\end{abstract}

\section{Introduction} \label{S:intro}

Many economic transactions are initiated by the buyer, and at the first possibility of contracting, the seller often does not yet know all the details, such as production costs. In this paper, we ask whether it is better for the seller to contract early or to wait until all relevant information is available. For example, consider a bilateral trade between a customer and a monopolistic service provider, such as a plumber or electrician. The buyer knows privately their willingness to pay for the repair and approaches the seller to get the job done. At this point, the seller does not yet know what needs to be done to fix the problem. One natural solution is to inspect the issue first and then, once the details are clear, provide a quote that the buyer can approve or reject. This mechanism has one weakness---since the work cannot begin until the uncertainty is resolved--- trade is delayed. Another approach is to offer a contract where the buyer agrees to cover the contractor's costs and fees, whatever they turn out to be, up to a specified limit, allowing work to start immediately. This type of contract avoids delays but creates potential incentive issues for the seller---after learning what the buyer is willing to pay, the seller may be incentivized to inflate the reported costs.

We model such situations using a standard bilateral trade framework, with three key assumptions: First, delaying the contract until the seller's private information is revealed incurs costs for both parties due to delay. Second, the seller cannot commit to truthfully reporting her private information. Third, the seller cannot ignore the information provided by the buyer. In scenarios where these assumptions do not hold, the optimal mechanism is straightforward: a take-it-or-leave-it offer based on the seller's type. However, under the conditions we consider, this proposed mechanism is not implementable.\footnote{If there were no cost of delay, both parties could simply wait until the seller's type is realized. If the seller could commit to truthfully revealing her private information once it is known, the optimal mechanism could be replicated with a type-contingent menu of mechanisms. Finally, if the seller could ignore the buyer's information, she would have no incentive to misrepresent her type once it is realized.}

The trade-offs we study here are relevant in significant public policy contexts. For instance, consider a pharmaceutical company acting as a monopolistic seller developing a new cure for some disease. In this scenario, a national health authority (the buyer) privately knows the value of this product for its population, but the seller is initially uncertain about the production costs, which will only become clear after extensive research. Delaying the contract until this information is available can be costly because early contracting would enable the health authority to organize approval and delivery much sooner. Although this paper focuses on cases where the principal is a seller and the agent is a buyer, our results also apply in reverse situations, provided the principal is the party that learns their private information with a delay. For example, consider a defense department as the buyer, contemplating the purchase of a new weapons system from a seller who privately knows the production costs. The defense department faces uncertainty regarding the future usefulness of the system. Over time, this uncertainty may resolve, but waiting for this information could result in delayed delivery of the system.

Our first main result shows that an optimal mechanism can effectively address the challenges posed by asymmetric information and timing in bilateral trade with information that arrives over time. Specifically, we show that a menu of two-part tariffs is an optimal mechanism under our assumptions. These two-part tariffs consist of a fixed fee and an \emph{at-will} delivery price. The buyer chooses a tariff, and the fixed fee is paid upfront. The seller commits to stick to the delivery price chosen by the buyer but reserves the right to refuse to trade upon learning her own type. In equilibrium, the seller will agree to trade when the price component of the tariff chosen by the buyer turns out to be higher than the seller's type, ensuring that the seller has no incentive to misreport her type. At the same time, it also implies that higher prices come with a higher probability of trade, which creates a potential for screening. However, if the menu only consisted of delivery prices, the buyer would not want to choose optimal prices from the seller's perspective. Thus, the tariff adds a specific fixed fee to each price, which can be positive or negative, and this fee adjusts expected payments to restore incentive compatibility for the buyer and optimality for the seller. In equilibrium, intermediate buyer types choose tariffs where the price is low, resulting in ex-post information rents that are too high, necessitating an additional charge to mitigate them. Conversely, the seller wants high types to choose very high delivery prices, which requires an ex-ante subsidy to provide the necessary informational rents. This combination leads to a novel form of non-monotonicity in price discrimination because as the delivery fee (i.e., per-unit price) increases, the corresponding fixed fee initially rises and then falls. This finding contrasts sharply with typical nonlinear pricing models, where a higher quantity (or probability of trade) is generally associated with a higher fixed fee.

Our second main result studies price mechanisms that do not allow the use of two-part tariffs. This analysis is motivated by the observation that optimal two-part tariff menus, especially those with the non-monotone structure we discussed, are rarely observed in practice. We begin by examining optimal static price mechanisms, where contracting occurs either before or after type information becomes available, without dynamic screening. When the discount factor is sufficiently high, the optimal static price mechanism involves waiting until the seller's type is revealed and then offering a take-it-or-leave-it price that will depend on that type. However, when the discount factor is low, the optimal price mechanism consists of a menu of offers made before the seller's type is revealed. This optimal menu consists of (possibly many) at-will prices and a guaranteed-delivery price. In equilibrium, the highest buyer types choose guaranteed delivery at a high price, while the at-will menu allows for further screening of intermediate buyer types. While these are the optimal static price contracts, we also show that when the discount factor is large enough, the optimal price mechanism becomes dynamic, offering early delivery to high types to minimize delay costs and late delivery to low types to mitigate incentive problems.

In the final part of our paper, we extend our model in two directions. First, we consider the possibility that the seller only learns her type if she pays a cost. This could occur, for example, if the good has never been produced before, requiring the seller to invest in order to learn the production cost, or if the good is available but the seller needs to perform research or hire an outside expert to evaluate its worth. We show that our main result still applies: if learning the cost is necessary in order to trade, the seller either chooses not to trade (if the cost is too high) or uses the two-part tariff mechanism but trades with a smaller set of buyers. If learning the cost is optional, she will offer to trade with guaranteed delivery when the cost is high. However, if the cost is sufficiently low, she will learn her type when the buyer type is intermediate and offer a two-part tariff, whereas if the buyer type is sufficiently high, she will offer a guaranteed delivery option. These high buyer types are willing to pay enough, so learning her type is not worth it for the seller.

Finally, we ask whether ex-post efficiency is implementable in our setting. In comparison with the classic \cite{myerson_efficient_1983} setting, we have the same constraints for the buyer, while the seller faces weaker participation constraints (she learns her type with delay) but stronger incentive compatibility constraints (she can update her information about the buyer's type before revealing her own). We show that in this setting, the former effect more than compensates for the latter, and ex-post efficiency is implementable. This is again achieved with two-part tariffs, but this time the seller always subsidizes the buyer regardless of whether trade occurs. 

\paragraph{Related literature.} The paper builds on the literature on seller profit maximization in mechanism design, starting with \cite{myerson_optimal_1981} and \cite{riley_optimal_1981}. Here, specifically, we look at the questions in the context of bilateral bargaining.  \cite{williams_efficient_1987} was the first to look at profit maximization in a bilateral bargaining context where both sides face Bayesian incentive compatibility and interim participation constraints. It found that the optimal allocation required that the seller's type be smaller than the buyer's virtual valuation. In this paper, we show that if the seller learns her type after the buyer, then that same allocation is an upper bound on what the seller can achieve in our context. More interestingly, we show that the same allocation can be obtained, but two-part tariffs are required. \cite{myerson_efficient_1983} is the seminal work in bilateral bargaining and shows that in the same environment as in \cite{williams_efficient_1987} ex-post efficiency cannot be implemented. In one of our extensions, we show that in our setting two-part tariffs allow for the implementation of ex-post efficient trade. 

The sequential way buyer and seller learn their type also introduces the possibility of considering both static and dynamic mechanisms. Dynamic contracts may be optimal if buyers arrive over time \citep{gershkov_dynamic_2009,board_revenue_2016} or buyers' valuations change or buyers learn about their values over time \cite{baron_regulation_1984,courty_sequential_2000,battaglini_long-term_2005,eso_optimal_2007,board_durable-goods_2008,pavan_dynamic_2014,garrett_intertemporal_2016,ely_overbooking_2017}. Importantly for our results, starting from \cite{stokey_intertemporal_1979,conlisk_cyclic_1984} this literature has emphasized that if buyers' valuations are persistent over time, there is no benefit in delaying the trade, i.e., intertemporal price discrimination does not help a monopolist. Our work differs from this literature in two ways: it is the principal who learns about her type over time and she does so only once (making it, by definition, persistent). In such situations however, we show that if we focus on price mechanisms and the discount factor is sufficiently high, dynamic contracting can be optimal. 

The fact that the principal privately learns over time connects our setting to informed principal problems, but with some crucial differences. The literature on the informed principal problem \citep{myerson_mechanism_1983,maskin_principal-agent_1990,maskin_principal-agent_1992,skreta_informed_2011,mylovanov_informed-principal_2012,mylovanov_mechanism_2014} assumes that the principal has private information at the contract design stage and cannot commit not to use this information. For the specific bilateral bargaining setting we are considering, however, \cite{yilankaya_note_1999} shows that the \cite{williams_efficient_1987} result, which assumes that mechanism is designed by a third party that wishes to maximize the seller's profit continues to hold when the informed seller designs the mechanism. It could be argued that we expand on this literature by looking at a ``not-yet informed'' principal problem, where the issue is not that the principal's choice of a mechanism itself may leak information to the agent but that the principal may not be relied on to report her private information, once learned, truthfully.

There is also a connection to the literature on limited commitment and Coasian dynamics. Since \cite{coase_durability_1972} and \cite{bulow_durable-goods_1982}, it has been known that a monopolist who lacks commitment power may not be able to exploit his position, as forward-looking buyers would simply wait and get a better price in the future. The literature on mechanism design with limited commitment has taken this as a starting point and studied various ways how the principal can at least partially overcome this problem 
\cite{bester_contracting_2001,vartiainen_auction_2013,gerardi_role_2014,deb_dynamic_2015,doval_optimal_2024,fugger_sequential_2019,liu_auctions_2019,doval_mechanism_2022}. In these models, the optimal contract is typically dynamic with a decreasing sequence of offers. In our model, with two-part tariffs the seller can achieve optimality by trading immediately, but when one focuses on pricing contracts, then as long as the buyer and seller value the future enough, dynamic contracts can improve on static ones. This is because a dynamic contract allows the seller to discriminate between high types---who trade immediately---and intermediate types, who trade later, once the seller's cost is known. The key here is that the seller does not have to learn exactly the buyer's type at time 0, only whether this type is high or not. This allows the seller to dampen her future's self incentives when trading with intermediate types.\footnote{There is also a connection to renegotiation proof contracts \citep{hart_contract_1988,dewatripont_commitment_1988}, which is another form of limited commitment: principals lack the ability to commit to a contract that may be ex-post Pareto dominated.}

Closely related to this point about commitment, there is also a connection to the literature on credibility in mechanisms \citep{akbarpour_credible_2020}. Just as in that literature, the principal can commit to the outcome function, but she cannot commit to revealing her private information truthfully. The difference is that in our setting, such information is her future type, whereas credibility concerns itself with the possibility that the principal may lie about the reports that participants in the mechanism may have sent her. 

Perhaps the closest paper to ours in terms of application is \cite{schmitz_how_2022}, which studies a similar bilateral trade problem where a buyer wants to procure a novel object, but the seller's costs are only revealed with delay. He shows that an at-will contract may generate more social surplus than a guaranteed delivery contract. The crucial difference is that in our setting it is the principal whose private information is learned with delay, whereas in \cite{schmitz_how_2022}, it is the agent's. The implications are very different: we show that the seller's incentive issues force the optimal contract to be a two-part tariff. If we limit ourselves to pricing, then optimal ex-ante price contracts always involve menus of offers, which have both an at-will and a guaranteed delivery component. This is also true in dynamic contracts.

The rest of the paper is as follows. In \cref{S:model}, we introduce the model, interpret it, and show how it can be reduced to a simpler model. In \cref{S:general}, we study general mechanisms, starting from benchmarks that provide an upper bound for our analysis, and then show how these outcomes could be achieved in our setting. In \cref{S:price}, we study a special class of simpler mechanisms we call price mechanisms. We first characterize the optimal static price mechanisms and then show that in some situations dynamic mechanisms can do strictly better than any static mechanism. In \cref{S: extensions}, we study two types of extensions to our analysis: costly learning and efficiency. Finally, \cref{S:discussion} concludes. In the \cref{S:proofs}, we provide proofs and additional technical details skipped in the text.

\section{Model} \label{S:model}

\subsection{Setup} \label{SS:setup}

\paragraph{Environment.} We consider a standard bilateral trade problem, where the seller has a single unit of an indivisible good available for sale to a buyer.\footnote{It is equivalent to a procurement problem with appropriate relabeling.} The buyer has a private type $\theta \in \Theta=[0,1]$, which represents his valuation for the object, while the seller has a private type $\omega \in \Omega=[0,1]$ that represents her valuation of the object.\footnote{Our results can be (with appropriate adjustments) generalized to a setting where $\theta$ and $\omega$ are distributed over arbitrary intervals, as long there is a non-empty intersection between such intervals. Using the $[0,1]$ interval simplifies the exposition.} There are two time periods $\tau \in \{0,1\}$: trade can be agreed at time 0 or at time 1, but if it is agreed at time 1, payoffs for both players have a common discount factor $\delta \in (0,1)$. 

We assume that it is common knowledge that $\theta$ has a distribution $F$ with density $f(\theta)>0$ while $\omega$ has a distribution $G$ with density $g(\omega)>0$. We assume these distributions are independent and satisfy the usual regularity conditions for both distributions. In particular, the regularity conditions guarantee that both the virtual valuation $\psi(\theta)=\theta - \frac{1-F(\theta)}{f(\theta)}$ and the virtual cost $\phi(\omega) = \omega + \frac{G(\omega)}{g(\omega)}$ are strictly increasing. For some of our results, we further assume that these are differentiable functions. As is standard, the notation $\one[A]$ represents the indicator function that is equal to 1 if A is true and 0 otherwise. Further, we will often use the notation $\gG(x)$ as a shorthand for $\int_0^x G(y)dy$ and the notation $\psi^+(\theta)$ as a shorthand for $\max[\psi(\theta),0]$.

\paragraph{Timing.} The timing unfolds as follows. To begin with, the seller commits to a mechanism $\gM = (\gB_0, \gB_1, \gS, q_0,q_1,t_0,t_1)$, which we describe below. 

At time 0, the buyer privately learns his valuation and chooses whether to participate in mechanism $\gM$. If the buyer chooses not to participate, both agents get a payoff $0$. If the buyer participates, he sends a public and contractible message $b_0 \in \gB_0$. The seller observes this message and privately learns the cost $\omega$. Only after this, the seller sends a public and contractible message $s \in \gS$. Then, the seller delivers the goods to the buyer with probability $q_0(b_0,s)$ and the buyer pays $t_0(b_0 s)$ to the seller.\footnote{Transfers could be negative, which, of course, means that the buyer receives a payment from the seller.}

At time 1, having observed $b_0$ and $s$, the buyer sends another public and contractible message $b_1 \in \gB_1$. Then, the seller delivers the good to the buyer with probability $q_1(b_0,b_1,s)$, and the buyer pays $t_1(b_0,b_1,s)$ to the seller. 

\paragraph{Payoffs.}
The buyer's utility after messages $(b_0,b_1,s)$ and with valuation $\theta$ is 
\[
  u(b_0,b_1,s|\theta) 
= \theta q(b_0,b_1,s)-t(b_0,b_1,s),
\]
where $q(b_0,b_1,s) = q_0(b_0,s) + \delta q_1(b_0,b_1,s)$ is the discounted allocation rule and $t(b_0,b_1,s) = t_0(b_0,s) + \delta t_1(b_0,b_1,s)$ is the discounted transfer rule. Similarly, the seller's profit is
\[
  v(b_0,b_1,s|\omega) 
= t(b_0,b_1,s)-\omega q(b_0,b_1,s).
\]

\paragraph{Strategies and beliefs.} A (behavioral mixed) strategy profile consists of $(\alpha,\beta_0,\sigma,\beta_1)$, where $\alpha(\theta)$ is the probability that the buyer participates in the mechanism, and (conditional on buyer participation) $\beta_0(\theta)$ is the probability distribution over the buyer's message space $\gB_0$ at time 0, $\sigma(\omega,b_0)$ is the probability distribution over the seller's message space $\gS$ at time 0, and $\beta_1(\theta,b_0,s)$ is the probability distribution over the buyer's message space $\gB_1$ at time 1. 

The only relevant belief updating happens when the seller observes $b_0$ and forms an updated belief about the buyer's type $\theta$. This is because at time 0, when the buyer makes his decisions, the mechanism has revealed no information about the seller's type, so the beliefs are equal to posterior $G$, and at time 1 when the buyer chooses the second message to send, the information about seller's type is irrelevant, as the buyer's payoff depends only on message $s$ and not the seller's type $\omega$ directly. Given this, with a slight abuse of notation, we denote the seller's updated beliefs by $F(\theta|b_0)$.

\paragraph{Equilibrium conditions.} We are looking for a perfect Bayesian equilibrium (PBE) of this game. That is, both buyer and seller behave optimally at each decision node, conditional on messages exchanged up to that point, and beliefs are updated according to Bayes' rule whenever possible. A pure strategy PBE has the following four equilibrium conditions\footnote{Mixed strategy PBEs are defined analogously, but the notation is somewhat more cumbersome, which is why we postpone these to \cref{A:revelationprinciple}.}
\begin{align}
  \int_0^1
    & u\left( 
      \beta_0(\theta), 
      \beta_1(\theta, \beta_0(\theta),\sigma(\omega,\beta_0(\theta))), 
      \sigma(\omega,\beta_0(\theta)) 
      | \theta \right)
    dG(\omega) \geq 0, \; \forall \theta \tag{IRB\textsubscript{0}} \label{E:IRBo} \\
    \beta_0(\theta) &\in \argmax_{b_0 \in \mathcal{B}_0} \int_0^1 u\left( b_0, \beta_1(\theta, b_0,\sigma(\omega,b_0)), \sigma(\omega,b_0) | \theta \right) dG(\omega), \;\;\; \forall \theta \tag{ICB\textsubscript{0}} \label{E:ICBo}
    \\
  \sigma(\omega,b_0) &\in \argmax_{s \in \mathcal{S}} \int_0^1 
    v\left( b_0, \beta_1(\theta, b_0,s), s | \omega \right)  
    dF(\theta|b_0), \;\;\; \forall \omega, \forall b_0 \in \mathcal{B}_0 \tag{ICS\textsubscript{0}} \label{E:ICSo}
  \\
  \beta_1(\theta,b_0,s) &\in \argmax_{b_1 \in \mathcal{B}_1} u\left( b_0, b_1, s | \theta \right),
  \;\;\; \forall \theta, \forall b_0 \in \mathcal{B}_0, \forall s \in \mathcal{S} \tag{ICB\textsubscript{1}} \label{E:ICBi}
\end{align}  
Additionally, we need to impose the feasibility constraints:
\begin{equation} \tag{F} \label{E:F}
  q_0(b_0,s), 
  q_1(b_0,b_1,s) \geq 0,
  q_0(b_0,s)+q_1(b_0,b_1,s) \leq 1, \;\;
  \forall b_0 \in \gB_0, b_1 \in \gB_1, s \in \gS
\end{equation}

\paragraph{Maximization problem.}
The seller's goal is to maximize her expected ex-ante profit (from now on, just profit) by choosing a mechanism $\gM$. That is, the maximization problem is\footnote{Again, we write it for the pure strategies here and postpone the mixed strategy formulation to \cref{A:revelationprinciple}.}
\begin{equation} \label{E:maxproblem}
  \max_{\gM} \int_0^1 \int_0^1 
    v(\beta_0(\theta),
      \beta_1(\theta,\beta_0(\theta),\sigma(\omega,\beta_0(\theta))),\sigma(\omega,\beta_0(\theta))|\omega) 
     dG(\omega)dF(\theta).
\end{equation}  
subject to \eqref{E:F}, \eqref{E:IRBo}, \eqref{E:ICBo}, \eqref{E:ICSo}, and \eqref{E:ICBi}.

\subsection{Interpretation} \label{SS:interpretation}

The model captures the dynamic interaction between the buyer and the seller without imposing additional commitment assumptions. First, consider a class of mechanisms where the seller opts not to engage with the buyer at the final stage (time 1). This is achieved by setting $q_1\equiv t_1 = 0$ and making $\gB_1 = \{0\}$. The transaction, if it happens, occurs at time 0. Any information the buyer communicates to the mechanism through messaging (message $b_0$) is received before the seller sends her message $s$. Thus, agreement must be reached before the seller's uncertainty is realized. Importantly, there is a separate set of \eqref{E:ICSo} constraints for each $b_0$, i.e., the mechanism must account for incentives of the seller, but there is no discounting.

Conversely, the seller can also choose to only trade at time 1 by setting $q_0 \equiv t_0 \equiv 0$ and $\gB_0 = \{0\}$. Here, at time 0, the seller learns the cost $\omega$ and sends the message $s$ without learning anything about the buyer's type. Thus, the set of \eqref{E:ICSo} constraints is reduced to one interim constraint, and the remaining constraints on the seller won't be binding.\footnote{This follows from \cite{williams_efficient_1987}. A detailed discussion is in \cref{SS:freelearning}} The issue is, however, that all payoffs in this scenario are discounted. 

\subsection{Reduced-Form Mechanisms} \label{SS:reducedform}

In many mechanism design problems, the revelation principle is crucial because it allows the analysis to focus only on direct mechanisms where agents report their types. In our model, the revelation principle does not apply because the information that the buyer reveals to the mechanism has a dual role: it enters the outcome function (as is standard), but it also may change the seller's beliefs and, hence, her incentives. Nevertheless, we can simplify the problem using a similar idea. Lemma \ref{T:revelationprinciple} shows that we can focus on reduced-form mechanisms, where both agents report only their types. To account for the dual role of information, we introduce a garbling device, which transfers information about the buyer's type to the seller and also influences early allocation.

Specifically, a reduced-form mechanism $\gM^R = \left( \gB, \mu, q_0,q_1,t_0,t_1 \right)$, is such that:
\begin{enumerate}
\item The buyer observes value $\theta \in \Theta$ and reports $\theta' \in \Theta$ to the garbling device.
\item The garbling device outputs a public message $b \in \gB$, with probability $\mu(b|\theta')$. This is common knowledge.\footnote{$\gB$ may be countable or uncountable so that $\mu(b|\theta)$ may be both a probability mass function and a (truncated) density.}
\item The seller observes message $b$ and upon learning the cost $\omega \in \Omega$, reports $\omega' \in \Omega$ to the mechanism.
\item Ex-ante transaction: with probability $q_0(b,\omega')$, the object is transferred to the buyer, and the latter pays $t_0(b,\omega')$ to the seller. 
\item Ex-post transaction: with probability $q_1(b,\omega',\theta')$ the object is transferred to the buyer, and the latter pays $t_1(b,\omega',\theta')$ to the seller. The payoffs from this transaction are discounted by $\delta \in (0,1)$.
\end{enumerate}

The seller's problem in the reduced-form environment is
\begin{equation} \label{E:reducedproblem}
   \max_{M \in \mathcal{M}^R} 
   \int_0^1
     \int_0^1  
     \E_{\mu(b|\theta)}
     [t(b,\omega,\theta)-\omega q(b,\omega,\theta)]
   dG(\omega)dF(\theta)
\end{equation}
 subject to $q_0,q_1 \in [0,1], q_0+q_1 \leq 1$,
 \begin{align}
   U(\theta|\theta) &\geq 0, \forall \theta,
   \tag{IRB} \label{E:IRB} \\
   U(\theta|\theta) &\geq U(\theta'|\theta) 
   , \forall \theta, \theta',
   \tag{ICB} \label{E:ICB} \\
   V(\omega|\omega,b)
   &
   \geq
   V(\omega'|\omega,b)
   ,
   \forall b,\omega, \omega',
   \tag{ICS}	\label{E:ICS}
 \end{align}
 where 
 \begin{align*}
    U(\theta'|\theta) 
    &= \int_0^1
        \E_{\mu(b|\theta')}
        \left[ 
          \theta q(b,\omega,\theta') - t(b,\omega,\theta')
        \right]
      dG(\omega)
      , \\
    q(b,\omega',\theta') 
    &= q_0(b,\omega') + \delta q_1(b,\omega',\theta'), \\
    t(b,\omega',\theta')
    &= t_0(b,\omega') + \delta t_1(b,\omega',\theta'), \\
    V(\omega'|\omega,b) 
    &= \int_0^1
     \left( t(b,\omega',\theta') - \omega q(b,\omega',\theta) \right)
    dF(\theta|b).
 \end{align*}
Remember that $F(\theta|b)$ is the posterior belief about the buyer's type $\theta$ after observing message $b$. Note also that we can assume without loss of generality that each $b \in \gB$ is meaningful, i.e., there exists $\theta$ such that $\mu(b|\theta)>0$. 

\begin{lemma}
\label{T:revelationprinciple}
For any mechanism $\gM$ and any equilibrium in it, there exists a reduced-form mechanism $\gM^R$ where a truthful equilibrium exists and guarantees both players the same equilibrium payoffs.
\end{lemma} 
The proof is in \cref{A:revelationprinciple}. Two remarks are in order. First, note that \Cref{T:revelationprinciple} does not say that for any mechanism and truthful equilibrium in the reduced problem \eqref{E:reducedproblem}, there exists a mechanism and an equilibrium in \eqref{E:maxproblem} that is payoff equivalent. This is because, in the reduced problem, the buyer only makes one announcement, leading to weaker constraints than in the general problem, where the buyer selects two distinct messages. However, this is still helpful if the mechanisms we identify through the reduced problem \eqref{E:reducedproblem} are feasible in the general problem \eqref{E:maxproblem} so that if they are optimal for the former, they will be optimal for the latter. Second, the reduced-form mechanism design problem includes a dynamic mechanism design problem (the choice of $q_0, q_1, t_0, t_1$) as well as an information design problem (the choice of $\gB$ and $\mu$). These are not, however, independent of each other as the choice of information structure $(\gB,\mu)$ affects the outcome functions.

\section{General Mechanisms}
\label{S:general}

In our reduced-form model, we make three crucial assumptions:
\begin{enumerate}
\item The seller cannot commit to revealing her type truthfully.
\item The garbling device's message is publicly observed.
\item If a trade occurs at time 1, both the buyer and seller's payoffs are discounted by $\delta \in (0,1)$.
\end{enumerate}
As implied by Lemma \ref{T:revelationprinciple}, these same assumptions hold for the general model.\footnote{In the general model, it is not the garbling device's message that is observed, but that of the buyer directly. The key point is that, in order to implement the mechanism, the seller must observe the variables that determine the outcome.}

In this section, we first consider changing the first assumption to a setting where the seller \emph{can commit} to revealing her type. This constitutes a natural benchmark and gives us an upper bound for the seller's profit among all possible mechanisms. We also show that this upper bound can be easily obtained if the seller could not observe any information from the garbling device and the implementation of the mechanism was delegated to a third party. Finally, we show that if $\delta = 1$, it would also be easy to achieve the upper bound by simply waiting for the seller to learn her cost and making trade possible only at time 1.

In \cref{SS:optimality}, we then show that, even when there is discounting and the seller cannot commit to both revealing her type and ignoring the information provided on the buyer's type, the benchmark allocation and profit for the seller can still be achieved. This is done through a mechanism where the buyer chooses from a menu of two-part tariffs. For each element of the menu, the first payment is a price for the good with an ``at-will" clause: the seller will deliver the good if she later learns that her type is below such price. The second payment is either a fee (if positive) or a subsidy (if negative) and is given regardless of whether trade occurs. We discuss this mechanism and argue that price mechanisms, where money changes hands only if there is trade, may be more realistic.

Before we begin the analysis itself, we argue that it is without loss to focus on a special class of reduced-form mechanisms.
\begin{definition} 
\label{D:full_revelation}
In a fully-revealing mechanism, $\gB=\Theta$ and $\mu(\theta|\theta)=1$. 
\end{definition}
In other words, in a fully-revealing mechanism, the buyer publicly announces her type. It is easy to see that in the benchmark case, where the seller commits to reveal her type truthfully, the seller has no role to play and none of the issues that the sequential nature of our problem arise. Given that, we can use the revelation principle and just focus on direct mechanisms, which in this case are equivalent to fully-revealing mechanisms where trade can only happen at time 0: $q_1 = t_1 = 0$.

When we then turn to analyzing the case where the seller does play a role, the focus on fully revealing mechanisms is, in principle, with loss. However, \cref{P:optimal} shows that the benchmark utility and profits for the seller can be obtained via a fully-revealing mechanism.

As we shall see in \cref{S:price}, when we impose that mechanisms are prices, fully revealing mechanism can sometimes be improved upon.  

\subsection{Benchmarks} \label{SS:freelearning}

\paragraph{S commits to reveal her type truthfully.} As discussed, above, when the seller plays no role, then we can apply the revelation principle. Thus, the problem becomes:
\begin{equation} \label{E:benchmark}
\max_{q_0,t_0}
\int_0^1 \int_0^1 [ t_0(\theta,\omega)-\omega q_0(\theta,\omega)] dF(\theta) dG(\omega)
\end{equation}
subject to
\begin{align*}	
	 \int_0^1 [\theta q_0(\theta,\omega)-t_0(\theta,\omega)] dG(\omega) &\geq 0, \forall \theta
	 \\
	 \int_0^1 [\theta q_0(\theta,\omega)-t_0(\theta,\omega)] dG(\omega) 
	&\geq 
	 \int_0^1 [\theta q_0(\theta',\omega)-t_0(\theta',\omega)] dG(\omega)
	, \forall \theta, \theta'
\end{align*} 

We then have:
\begin{proposition}
    \label{P:benchmark}
    The solution to \ref{E:benchmark} is the allocation:
    \begin{align*}
   q_0(\theta,\omega) = \one[\omega \leq \psi(\theta)],
\end{align*}
which can be implemented with transfers
\begin{align*}
t_0(\theta,\omega) = \psi^{-1}(\omega) \one[\omega \leq \psi(\theta)],
\end{align*}
The expected utility for the seller from this solution is
\begin{equation*}
  V^B 
  = \int_0^1 \int_0^1 \left(
    \psi(\theta) - \omega  
  \right) \one\left[\psi(\theta) \geq \omega\right] dF(\theta) dG(\omega) = \int_{\psi^{-1}(0)}^1\gG(\psi(\theta)) dF(\theta)
\end{equation*}	
\end{proposition}
The proof of this result is standard and, in fact, the result itself is implicit in \cite{williams_efficient_1987} as discussed below. We state it here explicitly for clarity. The proof is relegated to \cref{SS:proof_benchmark}. It is also immediate to check that this mechanism is also feasible in the general environment.

We refer to the allocation obtained here as the \emph{benchmark} allocation and to its corresponding expected utility for the seller as the \emph{benchmark} profit for the seller. One thing we wish to emphasize at this point is that while the transfers $t_0(\theta,\omega)$ described above are optimal, they are not unique.\footnote{This is because the buyer's constraints only put restrictions on the buyer's expected transfers.} Still, it is important to note that in the benchmark case, a natural implementation of the benchmark mechanism is one where the seller offers the buyer a price menu $\{\psi^{-1}(\omega)\}_{\omega \in \Omega}$ before $\omega$ is revealed. The buyer reports the highest price $p$ she is willing to accept; once $\omega$ is revealed, trade occurs at a price $\psi^{-1}(\omega)$ if this is no greater than $p$, while no trade or payments are made otherwise.

This is clearly not implementable if the seller cannot commit to truthfully revealing her type. Any maximal price $p$ chosen by the buyer effectively reveals her type to the seller. The latter, upon learning $\omega$, has an incentive, if $\psi^{-1}(\omega) < p$, to report instead that her type is $\omega'$, where $\psi^{-1}(\omega') = p$. We discuss below how our assumptions rule out two immediate solutions to this problem: the problem would not exist if the seller could commit to ignoring the information about the buyer she learns or if there was no discounting, allowing the seller to simply wait for her type $\omega$ to be revealed and then propose a single price $\psi^{-1}(\omega)$.

\paragraph{S does not observe the garbling device's message.} As we've shown above, removing incentive compatibility constraints for the seller leads to an upper bound for our problem. Now, we show that this upper bound could be reached if the seller did not observe the information that the garbling device provides and implementation was delegated to a third party. Formally, we are assuming here that problem \eqref{E:reducedproblem} holds in its entirety, but we impose that for any $b$, $F(\theta|b) = F(\theta)$. 
It is easy to see that with this restriction, the seller's expected profit when reporting $\omega'$ does not depend on $b$:
\begin{equation*}
    \int_0^1
        \E_{\mu(b|\theta)}
        \left[ 
          t(b,\omega',\theta) - \omega q(b,\omega',\theta)
        \right]
      dF(\theta)
\end{equation*}
and so, slightly abusing notation, we can rewrite (\ref{E:ICS}) as
\begin{equation}
  \int_0^1 
    [\Tilde{t}(\omega,\theta) - \omega \Tilde{q}(\omega,\theta)] dF(\theta) \geq  \int_0^1 
    [\Tilde{t}(\omega',\theta) - \omega \Tilde{q}(\omega',\theta)] dF(\theta), \forall \omega, \omega'.
    \tag{ICS$^\varnothing$}	\label{E:ICSR2}
\end{equation}
where $\Tilde{q}(\theta,\omega)=\mathbb{E}_{\mu(b|\theta)}q(b,\omega,\theta)$ and analogously for $\Tilde{t}$. But then we can apply to the constraint above, to the buyer's constraints, and to the objective function, the same construction just as we did in the case where the seller can commit to the truth. Hence, we end up with problem \eqref{E:benchmark} plus the incentive compatibility constraint for the seller \eqref{E:ICSR2} described above where $\Tilde{q}$ and $\Tilde{t}$ are replaced by $q_0$ and $t_0$. 

\cite{williams_efficient_1987} shows that the benchmark solution satisfies the \eqref{E:ICSR2} constraint, making $V^B$ achievable.\footnote{In fact, \cite{williams_efficient_1987} shows that we obtain the same solution if we impose \eqref{E:ICSR2} \emph{and} interim individual rationality for the seller, so Proposition 1 implies that these constraints are, in fact, slack when we are maximizing the seller's expected utility.} Thus, it is not the existence of an incentive compatibility constraint for the seller, per se, that is problematic. Rather, the problem arises if the seller announces her private information after updating her beliefs about the buyer's valuation.

Thus, a solution to the implementation of the benchmark mechanism would be a device that effectively implies that buyer and seller could agree to trade at time 0 without the latter knowing it.\footnote{As our examples in the introduction show, this is not a credible possibility in many realistic cases.}

\paragraph{No discounting.}
Suppose that $\delta=1$, so that waiting for $\omega$ to realize is without loss for either player. We now show that the benchmark payoff can be achieved. 
Since \[ q_0(b,\omega)+q_1(b,\omega,\theta) = q(b,\omega,\theta) \] 
(and, analogously for $t$), then we can wlog focus on mechanisms where $q_0 = t_0 = 0$ because $(q_1,t_1)$ can implement anything that $(q_0,t_0)$ can. 

Now, set $\gB = \{b\}$ so that it must be that $\mu(b|\theta) = 1$ for all $\theta$ and therefore $F(\theta|b)=F(\theta)$. Thus, effectively, $\gM^R = (q_1(\theta,\omega),t_1(\theta,\omega))$ since both $\gB$ and $\mu$ are degenerate and problem \eqref{E:reducedproblem} becomes again problem \eqref{E:benchmark} plus the \eqref{E:ICSR2} constraint, the only difference being that it applies to $(q_1,t_1)$ instead of $(q_0,t_0)$. Hence, the optimal mechanism is the benchmark mechanism where trade happens at time 1 and this is without loss as there is no discounting. 

When there is no discounting, the seller can, in essence, commit to not learning anything about the buyer's type by imposing that trade can only occur at time 1. Then, as we discussed in the previous case, she can implement the benchmark mechanism because it is incentive compatible for her future self (who still has prior beliefs on $\theta$) to truthfully report her type. Of course, this works, and it achieves the benchmark payoff for the seller, because there is no loss in waiting, but for any $\delta <1$, waiting would lead the seller to a strictly lower expected utility.

\subsection{Optimal Mechanisms}
\label{SS:optimality}

We now show that even without violating assumptions 1)-3), the benchmark allocation rule---and therefore the benchmark profit for the seller---can be achieved. This can be done in a fully-revealing mechanism with $q_1 = t_1 = 0$, as in the benchmark case, but cannot be done with a menu of prices.

Given that we are still looking at fully revealing mechanisms with $q_1=t_1=0$, but we are otherwise keeping all of our assumptions, the seller's problem \ref{E:reducedproblem} becomes 
\begin{equation}
\label{E:optimalproblem}
    \max_{q_0,t_0} 
   \int_0^1
     \int_0^1  
     [t_0(\theta,\omega)-\omega q_0(\theta,\omega)]
   dG(\omega)dF(\theta)
\end{equation}
 subject to $q_0\in [0,1]$,
 \begin{align*}
   \int_0^1
        \left[ 
          \theta q_0(\theta,\omega) - t_0(\theta,\omega)
        \right]
      dG(\omega)&\geq 0, \forall \theta
   \tag{IRB}\\
   \int_0^1
        \left[ 
          \theta q_0(\theta,\omega) - t_0(\theta,\omega)
        \right]
      dG(\omega)&\geq \int_0^1
        \left[ 
          \theta q_0(\theta',\omega) - t_0(\theta',\omega)
        \right]
      dG(\omega), \forall \theta,\theta'
   \tag{ICB} \\
       t_0(\theta,\omega) - \omega q_0(\theta,\omega) &\geq
       t_0(\theta,\omega') - \omega q_0(\theta,\omega'),
   \forall \theta,\omega, \omega',
   \tag{ICS}	
 \end{align*}

Problem \ref{E:optimalproblem} has a natural interpretation. The seller asks the buyer to report her type. Since the latter makes this report before observing the seller's report, she faces Bayesian IC and IR constraints. Having observed the buyer's type, and upon learning her own valuation, the seller reports. Since she knows the buyer's type, her IC constraint is in dominant strategies. Thus, the asymmetry in timing in our setting leads to a problem where the two sides' constraints have different informational requirements.\footnote{In certain settings Bayesian and dominant strategy implementation are outcome equivalent. We implicitly impose ex-post balancedness as is standard in bilateral bargaining problems, and it is well known (\cite{manelli_bayesian_2010}) that where ex-post balancedness is required, this outcome equivalence does not necessarily hold.} 

The following result describes our optimal mechanism:

\begin{proposition} \label{P:optimal}
The solution to problem \eqref{E:optimalproblem} is a mechanism that implements the benchmark allocation: 
	\begin{align*} 
 q_0(\theta,\omega) = \one[\omega \leq \psi(\theta)],
 \end{align*}
 with transfers
 \begin{align*}
 t_0(\theta,\omega) = \lambda(\theta) + \psi(\theta) \one[\omega \leq \psi(\theta)],
	\end{align*}
 where \begin{equation*}
     \lambda(\theta)= (\theta - \psi(\theta))G(\psi(\theta)) - \int_0^\theta G(\psi(y)) dy,
 \end{equation*} 
 whenever $\theta \geq \psi^{-1}(0)$ and $\lambda(\theta)=0$ otherwise.
\end{proposition}
 \begin{proof}
     As is standard, IC and IR for the buyer imply that $Q_0^B(\theta) = \int_0^1 q_0(\theta,\omega) dG(\omega)$ is weakly increasing and that
     \begin{align*}
         T_0^B(\theta)=\int_0^1 t_0(\theta,\omega) dG(\omega)=\theta Q_0^B(\theta)- \int_0^\theta Q_0^B(y)dy
     \end{align*}
     with $T^B(0)=0$. 
     
     Now, given $v(\theta,\omega)=t_0(\theta,\omega) - \omega q_0(\theta,\omega)$ then IC for the seller implies that $q_0(\theta,\omega)$ be weakly decreasing in $\omega$ for any $\theta$ and that
     \begin{align*}
         t_0(\theta,\omega)= v(\theta,1) + \int_\omega^1 q_0(\theta,x) dx + \omega q_0(\theta,\omega).
     \end{align*}
     We denote
\begin{equation*}
\lambda\left( \theta \right) =v\left(\theta,1 \right),
\end{equation*}%
then we have
\begin{equation*}
T_0^{S}\left( \theta \right) =\int_{0}^{1}t_0\left( \omega ,\theta \right)
dG\left( \omega \right) =\lambda\left( \theta \right) +\int_{0}^{1}\left[
\int_{\omega }^{1}q_0\left( x,\theta \right) dx+\omega q_0\left( \omega ,\theta
\right) \right] dG\left( \omega \right).
\end{equation*}%
Suppose $q_0\left( \omega ,\theta \right) =1\left[ \omega \leq \psi\left(
\theta \right) \right].$ We show below that this allocation is implementable in this problem and since it is the optimal allocation for the relaxed problem without seller constraints, it must also be the optimal allocation here. Recalling that $\psi^+(\theta)=\max\{\psi(\theta),0 \}$, 
we have 
\begin{align*}
T_0^{S}\left( \theta \right)
&=\lambda\left( \theta \right) +\int_{0}^{1}\left[
\int_{\omega }^{1}\one\left[ x\leq \psi\left( \theta \right) \right]
dx+\omega \one\left[ \omega \leq \psi\left( \theta \right) \right] \right]
dG\left( \omega \right)  \\
&=\lambda\left( \theta \right) +\int_{0}^{1}\one\left[ \omega \leq \psi\left(
\theta \right) \right] G\left( \omega \right) d\omega +\int_{0}^{1}\omega \one%
\left[ \omega \leq \psi\left( \theta \right) \right] dG\left( \omega
\right)  \\
&=\lambda\left( \theta \right) +\mathcal{G}\left( \psi ^{+}\left( \theta \right)
\right) +\psi ^{+}\left( \theta \right) G\left( \psi ^{+}\left( \theta
\right) \right) -\mathcal{G}\left( \psi ^{+}\left( \theta \right) \right)
\\
&=\lambda\left( \theta \right) +\psi ^{+}\left( \theta \right) G\left( \psi^{+}\left( \theta \right)\right) 
\end{align*}%
while%
\begin{align*}
T_0^{B}\left( \theta \right) 
&=\theta Q_0^{B}\left( \theta \right)
-\int_{0}^{\theta }Q_0^{B}\left( y\right) dy \\
&=\theta \int_{0}^{1}\one\left[ \omega \leq \psi\left( \theta \right) %
\right] dG\left( \omega \right) -\int_{0}^{\theta }\int_{0}^{1}\one\left[
\omega \leq \psi\left( y\right) \right] dG\left( \omega \right) dy \\
&=\theta G\left( \psi ^{+}\left( \theta \right) \right) -\int_{0}^{\theta
}G\left( \psi ^{+}\left( y\right) \right) dy.
\end{align*}%
Of course, we need $T_0^{S}\left( \theta \right) =T_0^{B}\left( \theta \right) $
which then implies%
\begin{gather*}
\lambda\left( \theta \right) =\left( \theta -\psi ^{+}\left(
\theta \right) \right) G\left( \psi ^{+}\left( \theta \right) \right)
-\int_{0}^{\theta }G\left( \psi ^{+}\left( y\right) \right) dy
\end{gather*}%
Given the definition of $\psi^+(\theta)$, this is the $\lambda(\theta)$ function we were looking for. It is immediate to check that $q_0\left( \omega ,\theta \right) =\one\left[ \omega \leq \psi\left(
\theta \right) \right]$ is weakly decreasing in $\omega$ for all $\theta$, that $Q_0^B(\theta)=G(\psi^+(\theta))$ is weakly increasing in $\theta$, and that type $\theta=0$ has expected utility equal to zero. Finally, 
\begin{align*}
         t_0(\theta,\omega)= \lambda(\theta) + \int_\omega^1 \one[x \leq \psi(\theta)]dx + \omega \one[\omega 
 \leq \psi(\theta)]=\lambda(\theta)+\psi(\theta) \one[\omega \leq \psi(\theta)]
     \end{align*}           
 \end{proof}

\Cref{P:optimal} tells us that the benchmark allocation is implementable even with the additional dominant-strategy IC constraint for the seller, but this requires very specific transfers, which require payments even when trade does not take place. The corollary below collects some properties of these payments:

\begin{corollary}
\label{C:properies_v}
Let $\theta^* = \psi^{-1}(0)$. Then $\lambda(\theta)$ is everywhere continuous and a differentiable function for $\theta > \theta^*$. Further: 
\begin{enumerate}
    \item $\int_0^1 \lambda(\theta) dF(\theta) = 0$,
    \item $\lambda_{\theta}^{'}(\theta^*)>0$,
    \item $\lambda(1) < 0$.
\end{enumerate}
\end{corollary}
The proof is in \Cref{A:properies_v}.
The first property confirms that this mechanism achieves the benchmark expected utility for the seller since the expected utility in this mechanism is
\begin{align*}
&\int_0^1 \int_0^1 [t_0(\theta,\omega) - \omega q_0(\theta,\omega)] dG(\omega) dF(\theta) 
\\
=&\int_0^1 \int_0^1 [\lambda(\theta)+(\psi(\theta) - \omega) \one[\psi(\theta) \geq \omega]] dG(\omega) dF(\theta)
\\
=&V^B + \int_0^1 \int_0^1\lambda(\theta) dG(\omega) dF(\theta) = V^B
\end{align*}
In the mechanism, buyer types for which the virtual valuation is non-positive never trade and don't pay any transfers. The second and third properties imply that buyers whose virtual value is just above zero, in equilibrium make a positive payment $\lambda(\theta)$ to the seller and agree to pay a further price $\psi(\theta)$ for the good if the seller agrees to deliver the good once her type is revealed. If the seller decides not to deliver the good, they don't pay this additional price. In equilibrium, the seller will agree to deliver the good if $\omega \leq \psi(\theta)$. Types who are sufficiently close to 1, agree to the same deal, \emph{but} their $\lambda(\theta)$ is negative so that for these types, $\lambda(\theta)$ represents a subsidy they receive from the seller regardless of whether the latter delivers or not.\footnote{When $F$ and $G$ are uniform distributions, we can show that there exists a $\theta^{**}$ such that $\lambda(\theta)>0$ for all $\theta \in (\theta^*,\theta^{**})$ and $\lambda(\theta) < 0$ for all $\theta \in (\theta^{**},1]$ but for generic distribution, we can only characterize $\lambda(\theta)$'s behavior in the neighborhoods of $\theta^*$ and 1.}

To get an intuition for this mechanism, note that the transfers we had in the benchmark mechanism are a function of the seller's type. In a setting where the seller cannot commit to truthfully revealing her type, such a mechanism will not be implementable as the seller will sometimes have the incentive to misreport her type to achieve a better price. So, the mechanism makes the transfers a function of the buyer's type instead. In particular, the seller proposes to the buyer that he chooses a price $p$ he is willing to pay for the good, with the proviso that the seller will only deliver the good if her type turns out to be no greater than $p$. This is clearly incentive compatible for the seller. In addition, if each buyer type $\theta$ chooses $p(\theta) = \psi(\theta)$ whenever $\psi(\theta) \geq 0$ and proposes no price otherwise, then we would implement the benchmark allocation. The problem is that $p(\theta) = \psi(\theta)$ is not incentive compatible for the buyer. For example, since $\psi(1) = 1$, type $\theta = 1$ would commit to pay 1 for a certain purchase of a good that is worth 1 to him, resulting in a total utility of zero. He is clearly better off choosing a price $p < 1$ for an expected utility of $(1 - p)G(p) > 0$.

If we allowed buyer types to choose prices according to an incentive-compatible price schedule $p(\theta) \neq \psi(\theta)$, but still insisted that the seller would deliver the good only if $\psi(\theta) \geq \omega$ we'd by definition achieve incentive compatibility for the buyer, and this would respect the benchmark allocation. But it's easy to see that we'd lose again incentive compatibility for the seller. This is because whenever $p(\theta) > \omega > \psi(\theta)$ the seller would have an incentive to underreport her type and make the trade happen and, conversely, whenever $\psi(\theta) > \omega > p(\theta)$, the seller would have an incentive to overreport her type and stop the trade. 

What the optimal mechanism does is to fix the problem by making $p(\theta) = \psi(\theta)$ so that we have incentive compatibility for the seller and implementation of the benchmark allocation, but also adds a payment $\lambda(\theta)$ that is independent of the allocation but changes the total net payments that the buyer has to pay in order to achieve incentive compatibility for him. Going back to the $\theta=1$ example, the mechanism asks this buyer type to pay a price of 1 in exchange for a certain delivery of a good that he values 1, but as the third property of Corollary \ref{C:properies_v} emphasizes, he also get a subsidy from the seller to compensate. In equilibrium, therefore, type 1 gets no utility from trade as such, but her informational rents are given by this subsidy. For other types, making it incentive compatible for a certain type to choose a price equal to his virtual valuation requires a smaller subsidy or even the payment of a fee.\footnote{In other words, when $\lambda(\theta)>0$ the seller can effectively charge more than the virtual valuation and still preserve incentive compatibility with this buyer type, while when $\lambda(\theta) <0$, the virtual valuation is too high a price to be incentive compatible and the seller needs to subsidize the buyer.}

\begin{figure}[H]
    \centering
    \includegraphics[width=0.7\linewidth]{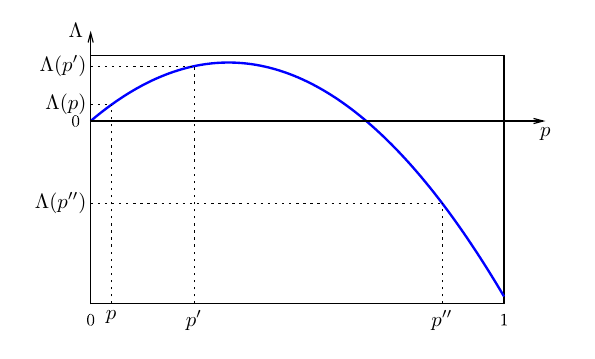}
    \caption{Optimal menu of two-part tariffs in the case of uniform distributions.}
    \label{F:lambda}
\end{figure}

The discussion above suggests an immediate, indirect implementation: the seller offers a menu of payment pairs $\{(p,\Lambda(p)\}_{p \in [0,1]}$, where $\Lambda(p)=\lambda(\psi^{-1}(p))$. \Cref{F:lambda} illustrates this menu in the case of uniform $F$ and $G$ distributions. Each option in the menu is an \emph{at-will} contract plus a fixed transfer. If the buyer chooses a particular price $p$, the buyer commits to paying that price, but the seller retains the right to not deliver upon learning her type. If she does not deliver, then $p$ is not paid. The buyer also accepts an up-front fee or subsidy equal to $\Lambda(p)$ which is paid whether trade occurs or not. This is therefore a menu of \emph{two-part tariffs}, with the unusual feature that for some components of the menu, the fixed component is a fee while for others it is a subsidy. 

The fact that this seems unusual suggests that there may be other factors that make it unpractical. It is simple to observe, for instance, that those buyer types who pay a fee may end up not getting the good which immediately implies that this mechanism is not ex-post individually rational for the buyer; this may be a concern in some environments. It may also be that the mechanism is too complex and relies excessively on fine details of the problem: the natural implementation of the benchmark mechanism only requires buyers to think about the most they are willing to pay, while this mechanism asks them to choose from a menu of complicated pricing options. For example, it may be that $p'' > p' > p$ but that $\Lambda(p') > \Lambda(p) > \Lambda(p'')$ and so a buyer would find it difficult to compare the three options.\footnote{Recall that we know that $\lambda(\theta)$ is positive and increasing in a neighborhood of $\theta^*$ and negative in a neighborhood of $\theta =1$ so it cannot be monotone.} \Cref{F:lambda} illustrates this comparison.

For this reason, it is worthwhile to look at a simpler class of mechanisms, one where buyers only make a payment if trade takes place. We study such \emph{price} mechanisms in the next section.

\section{Price Mechanisms} \label{S:price}

\subsection{Definitions}

In the previous section, we introduced the idea that a simple mechanism would have the property that buyers only make a payment if trade takes place. We begin by defining this notion formally. Next, we show that this property is not enough. In particular, we show in Example 1 that one can construct a mechanism that satisfies this property and ``almost'' implements the benchmark allocation, but this mechanism involves randomization and is significantly less realistic than even two-part tariffs. 

\begin{definition}
A reduced-form mechanism $\gM^R$ satisfies the \emph{no payment without trade} (NPWT) property if for all $b \in \gB, \theta \in \Theta,$ and $\omega \in \Omega$,
\[
 q_0(b,\omega) = 0 \Rightarrow t_0(b,\omega) = 0 \text{ and }
 q_1(b,\omega,\theta) = 0 \Rightarrow t_1(b,\omega,\theta) = 0.
\]
\end{definition}  
While the (NPWT) property eliminates the possibility of a literal use of two-part tariffs, it still allows for mechanisms where introducing randomization through the garbling device can approximately deliver the benchmark outcome.\footnote{A similar example can be constructed with a garbling device that is deterministic, but the allocation and transfer rules are stochastic.}

\paragraph{Example 1.} Consider a mechanism where the garbling device $(\gB,\mu)$ is such that $\gB = \Theta \cup { \varnothing }$, $\mu(\varnothing|\theta)=\varepsilon$, $\mu(\theta'|\theta)=1-\varepsilon$ when $\theta'=\theta$, and $\mu(\theta'|\theta)=0$ when $\theta' \in \Theta \setminus {\theta}$. Let the allocation rule be $q_0(\varnothing,\omega)=1$ and $q_0(\theta,\omega)=\one[\omega \leq \psi(\theta)]$ for all $\theta \in \Theta$, and $q_1(b,\omega,\theta) \equiv 0$. In other words, the message $\varnothing$ says “deliver the object with certainty,” and other messages lead to the benchmark allocation rule. Let $t_1(b,\omega,\theta) \equiv 0$, while $t_0(\theta,\omega) = \max\{0,\psi(\theta)\}$ and $t_0(\varnothing,\omega)=\lambda^{\varepsilon}(\theta) / \varepsilon$, where $\lambda^{\varepsilon}(\theta)$ is computed so that incentive compatibility for the buyer is satisfied, i.e.,
\[
  \lambda^{\varepsilon}(\theta)
 = 
 (1-\varepsilon) \left( 
    \theta G(\psi(\theta)) - \int_0^{\theta} G(\psi(x) dx 
 - 
    \int_0^{\psi(\theta)} \psi(\theta) dG(\omega)
 \right).
\]
This mechanism satisfies all constraints and the (NPWT) property by construction. Moreover, as $\varepsilon \to 0$, this mechanism converges to the benchmark mechanism and achieves profit arbitrarily close to it.

The mechanism achieves this by introducing a probability $\varepsilon$ that trade takes place regardless of the underlying types. It is easy to see that as this probability approaches zero, the payment required by the buyer $\lambda^{\varepsilon}(\theta) / \varepsilon$ becomes arbitrarily large. This is neither simpler nor more realistic than two-part tariffs. Therefore, we focus on a class of simpler deterministic mechanisms that satisfy the (NPWT) property.

\begin{definition}
A reduced-form mechanism $\gM^R$ is deterministic if
\begin{enumerate}
\item It has a deterministic garbling device, i.e., there exists a function $m : \Theta \to \gB$ such that $\mu(b|\theta)=\one[b=m(\theta)]$ for all $b \in \gB$, $\theta \in \Theta$, and
\item It has a deterministic allocation rule, i.e., $q_0(b,\omega) \in \{0,1\}$ and $q_1(b,\omega,\theta) \in \{0,1\}$ for all $b \in \gB$, $\theta \in \Theta$, $\omega \in \Omega$.
\end{enumerate}
\end{definition}
Note that with the deterministic garbling device, the first argument of the $q_1$ and $t_1$ functions is redundant. Therefore, with slight abuse of notation, we replace
$q_1(m(\theta),\omega,\theta)$ with $q_1(\theta,\omega)$ and $t_1(m(\theta),\omega,\theta)$ with $t_1(\theta,\omega)$.

\begin{definition}
A price mechanism is a deterministic reduced-form mechanism $\gM^R$ that satisfies the (NPWT) property. We denote with $\gM^P$ the class of such mechanisms.
\end{definition}

\subsection{Static Price Mechanisms} \label{SS:static}

Static price mechanisms are simple and realistic types of mechanisms that provide useful benchmarks. By a static mechanism, we mean that trade only happens at a specific time (if at all), so there is no room for dynamic pricing. Specifically, we use the following definitions:
\begin{definition}
A price mechanism is
\begin{enumerate}
\item An \emph{ex-post} price mechanism if $q_0(b,\omega)=t_0(b,\omega)=0$ for all $b \in \gB$, $\omega \in \Omega$. We denote with $\gM^{EPP} \subset \gM^P$ the class of ex-post price mechanisms.
\item An \emph{ex-ante} price mechanism if $q_1(\theta,\omega)=t_1(\theta,\omega)=0$ for all $b \in \gB$, $\theta \in \Theta$, $\omega \in \Omega$. We denote with $\gM^{EAP} \subset \gM^P$ the class of ex-ante price mechanisms.
\item A \emph{static} price mechanism if it is either an ex-post or ex-ante price mechanism.
\item A \emph{dynamic} price mechanism if it is a price mechanism that is not static.
\end{enumerate}
\end{definition}

Intuitively, ex-post price mechanisms are those where trade can only be agreed at time 1, after the cost uncertainty has already been realized, whereas ex-ante price mechanisms are those where trade can only be agreed at time 0, before the cost information is revealed. In this subsection, we characterize the optimal ex-post and ex-ante price mechanisms, and the one of the two that guarantees higher profit is the optimal static price mechanism.

\subsubsection{Ex-Post Price Mechanisms} \label{SSS:expost}

We first consider ex-post price mechanisms. In this class of mechanisms, the seller's maximization problem is
\begin{align}
\label{E:EPOproblem}
	\max_{\gM^{EPP}}
	\delta \int_0^1 \int_0^1 [ t_1(\theta,\omega)-\omega q_1(\theta,\omega)] dF(\theta) dG(\omega)	
\end{align}
	subject to
	\begin{align*}	
		\delta  \int_0^1 [\theta q_1(\theta,\omega)-t_1(\theta,\omega)] dG(\omega) &\geq 0, \forall \theta
		\\
		\delta \int_0^1 [\theta q_1(\theta,\omega)-t_1(\theta,\omega)] dG(\omega) 
		&\geq 
		\delta  \int_0^1 [\theta q_1(\theta,\omega)-t_1(\theta',\omega)] dG(\omega)
		, \forall \theta, \theta'
		\\
		\delta  \int_0^1 [t_1(\theta,\omega)- \omega q_1(\theta,\omega)] dF(\theta)
		&
		\geq
		\delta  \int_0^1 [t_1(\theta,\omega')- \omega q_1(\theta,\omega')] dF(\theta)
		,
		\forall \omega, \omega'		
\end{align*}
This is exactly the same problem as \eqref{E:benchmark} in \cref{SS:freelearning}, where the seller commits to ignoring the report by the buyer, with one crucial difference---everything happens at time 1 instead of time 0 and therefore all payoffs are multiplied by $\delta$. The solution to that problem was clearly a price mechanism. Therefore the optimal mechanism within this class, which we call ex-post optimal (EPO) price mechanism, is such that $q_1(\theta,\omega) = \one[\psi(\theta) \geq \omega]$, and $t_1(\theta,\omega) = \psi^{-1}(\omega) \one[\psi(\theta) \geq \omega]$.The only difference is that the profit is discounted by $\delta \in (0,1)$, i.e., it gives profit $V^{EPO} = \delta V^B$.

\subsubsection{Ex-Ante Price Mechanisms} \label{SSS:exante}

Let us now consider the class of ex-ante price mechanisms. The seller's maximization problem is now
\begin{equation} \label{E:exanteproblem} 
 \max_{\gM^{EAP}} \int_0^1 \int_0^1
 [t_0(m(\theta),\omega)-\omega q_0(m(\theta),\omega)]
 dG(\omega) dF(\theta)
\end{equation}
\begin{align}
\int_0^1 [\theta q_0(m(\theta),\omega)-t_0(m(\theta),\omega)] dG(\omega) &\geq 0, \forall \theta
 \notag
\\
\int_0^1 [\theta q_0(m(\theta),\omega)-t_0(m(\theta)\omega)] dG(\omega) 
&\geq \int_0^1 [\theta q_0(m(\theta'),\omega)-t_0(m(\theta'),\omega)] dG(\omega), \forall \theta, \theta'
\notag
\\
t_0(b,\omega)- \omega q_0(b,\omega)
&\geq
 t_0(b,\omega')- \omega q_0(b,\omega'), \forall b, \omega, \omega'
\notag
\end{align}
In problem (\ref{E:EPOproblem}), the seller only had prior information and, as discussed in \cref{SS:freelearning}, having no updated information slackens completely the seller's incentive compatibility problem. When we looked at general mechanisms, we also saw that if we allowed for two-part tariffs, then a fully-revealing mechanisms could still implement the benchmark outcome. Here, we are only looking at ex-ante price mechanisms so it is not obvious whether it is optimal for the seller to allow her future self to update her beliefs about the buyer's type or not. 

The answer is in the lemma below:
  \begin{lemma} \label{L:FREAisEAO}
 The solution to problem \eqref{E:exanteproblem} and the solution to problem \eqref{E:optimalproblem}---when this is limited to mechanisms in $\gM^{EAP}$---coincide.
\end{lemma} 
The proof is in \Cref{A:FREAisEAO}.

Thus, it turns out that within the class of ex-ante price mechanisms, fully revealing mechanisms are without loss of generality. This result follows entirely from the assumption that the garbling device is deterministic. It provides a helpful simplification, as it eliminates the information design aspect of the problem. 
 
\paragraph{Menus of offers.} It is useful to define a particular class of simple mechanisms. As discussed, an at-will offer is a price offer that gives the seller the opportunity to walk away if the cost is too high. In this particular context, it is a price $\widehat{p}_0$ such that if B accepts this, then the trade happens at price $\widehat{p}_0$ if and only if S learns that the cost $\omega \leq \widehat{p}_0$. Another type of offer is a guaranteed delivery price, where the seller delivers the object no matter the cost. A menu of offers is a collection of at-will and guaranteed delivery offers. Specifically, each menu of offers is described by a function $p_0 : \Theta \to \R$ and a constant $P_0 \geq 0$, such that each $p_0(\theta)$ is an at-will offer, and $P_0$ is the guaranteed delivery price.\footnote{The assumption that there is only one guaranteed delivery price is without loss of generality, as the buyer would choose only the lowest price.}
It is straightforward to verify that menus of offers can be implemented as ex-ante price mechanisms. The following establishes this optimality---a menu of offers that satisfies particular conditions is indeed a solution to problem \eqref{E:exanteproblem}.

\begin{proposition} \label{P:auxiliary_EAO}
Problem \eqref{E:exanteproblem} is solved by a menu of offers, described by $p_0(\theta) \in [0,\theta]$ and $P_0 \geq 0$, such that:
\begin{enumerate}
\item Function $p_0(\theta)$ is weakly increasing and satisfies the following equation for each $\theta \in \Theta$,
\begin{equation} \label{E:auxpzerocond}
	\left( \theta-p_0(\theta) \right) G(p_0(\theta))		
	= \int_0^{\theta} G(p_0(x)) dx.
\end{equation}
\item There exists a marginal buyer type $\thetaone < 1$ so that buyers with types $\theta > \thetaone$ take the guaranteed delivery option at price $P_0$.
\item The guaranteed delivery price satisfies
\begin{equation} \label{E:auxPcond}
	P_0 = p_0(\thetaone)  G(p_0(\thetaone)) + \thetaone [1-G(p_0(\thetaone))].
\end{equation}
\end{enumerate}	
\end{proposition}
The proof is in \cref{A:auxiliary_EAO}.

While the previous result does not fully determine the optimal mechanism, it puts a specific structure on it and within this class the problem can be solved numerically for specific distributions $F$ and $G$. The following corollary describes some properties of any weakly increasing $p_0(\theta)$ function that satisfies constraint \eqref{E:auxpzerocond}.	
\begin{corollary} \label{C:auxiliary_EAO}
	$p_0(\theta)$ from \cref{P:auxiliary_EAO} satisfies the following properties:	
	\begin{enumerate} 
	\item If $p_0(\theta)$ is differentiable at $\theta$ then either $p_0'(\theta) = 0$ or $p_0'(\theta) = \phi^{-1}(\theta)$, where $\phi(x)$ is the virtual cost.
	\item If there is a discontinuous jump at $\theta$ from $\up$ to $\op>\up$, then
	$\op < \phi^{-1}(\theta) < \up$ and 
	\[
		\frac{\op G(\op) - \up G(\up)}{G(\op) - G(\up)} = \theta.
	\]
	\end{enumerate}	
\end{corollary}

It is interesting to compare this with the case where both buyer and seller face Bayesian incentive compatibility or dominant strategy incentive compatibility constraints. In the case of Bayesian incentive compatibility, as previously discussed, the solution was originally provided by \cite{williams_efficient_1987} and it is the benchmark mechanism, where $q_0(\theta,\omega)=\one[\psi(\theta) \geq \omega]$ whereas here we have $q_0(\theta,\omega)=\one[p_0(\theta) \geq \omega]$ for all $\theta < \thetaone$ and $q_0(\theta,\omega) = 1$ for all $\theta \geq \thetaone$. Therefore it is clear that the ex-ante optimal price mechanism cannot achieve the benchmark profits. The following corollary formalizes this observation.

\begin{corollary} \label{C:EAOvsB}
  The profit from the optimal ex-ante price mechanism is strictly lower than $V^B$. 
\end{corollary}	
The proof is in \Cref{A:EAOvsB}.

On the other hand, when both buyer and seller face dominant-strategy incentive compatibility constraints, \cite{hagerty_robust_1987} showed that the optimal solution is an at-will offer, but the seller offers a unique such price that all buyer types must either accept or reject. The result above shows such a mechanism cannot be an optimal ex-ante price mechanism because it does not feature a guaranteed delivery offer.

It is natural to ask, whether the optimal ex-ante price mechanism can sometimes be a single guaranteed-delivery offer with no at-will price $p_0(\theta)$. The answer is no. \begin{corollary} \label{C:EAOvsEASI}
  The profit from the optimal ex-ante price mechanism is strictly higher than the profit obtainable from a mechanism in which the buyer is offered only a guaranteed delivery price $P_0$. 
\end{corollary}
The proof is in \Cref{A:EAOvsEASI}. In it, we show that a single guaranteed delivery offer would give the seller a profit that would be strictly lower than the profit from a single at-will offer (with an optimally chosen price). In turn, \Cref{P:auxiliary_EAO} shows that such a single at-will offer is strictly dominated by a menu of offers, which includes a guaranteed delivery offer. Therefore, the optimal ex-ante price mechanism is always a menu of offers, which features at least one at-will offer and one guaranteed delivery offer.

\subsubsection{Optimal Static Mechanisms}

Combining the results from the previous two subsections, we can now state the following corollary from previous results.
\begin{corollary} \label{C:optimal_static}
There exists $\deltastatic \in (0,1)$ such that the optimal static price mechanism is
\begin{enumerate}
\item the EAO mechanism with profit $V^{EAO}<V^B$ when $\delta \leq \deltastatic$,
\item the EPO mechanism with profit $V^{EPO}=\delta V^B$ when $\delta \geq \deltastatic$.
\end{enumerate}
\end{corollary}
The proof is follows directly from the results above.

The following figure summarizes the results from this section when both $F$ and $G$ are uniform distributions. The green line depicts the benchmark profit $V^B$, which is the highest possible profit when the seller can use arbitrary mechanisms, including two-part tariffs. The increasing black line shows $V^{EPO} = \delta V^B$, which is the profit from the optimal ex-post mechanism (which turns out to be a price mechanism). Clearly, as $\delta \to 1$, it converges to $V^B$, but for any $\delta < 1$, the profit is strictly lower. The blue solid line shows the optimal ex-ante price mechanism discussed above. It is a combination of at-will and guaranteed delivery offers. The profit from ex-ante mechanisms does not depend on the discount factor $\delta$. Therefore, naturally, if the discount factor is low, $V^{EAO} > V^{EPO}$, and if the discount factor is close to one, then $V^{EAO} < V^{EPO}$. The maximum of these two profits is the profit from the optimal static mechanism, $V^{static} = \max\{V^{EAO}, V^{EPO}\}$. Finally, the figure also depicts two other natural mechanisms. The purple line shows the highest profit achievable by a single at-will offer, which turns out to be the best the seller can do using a menu of at-will offers. \Cref{P:auxiliary_EAO} shows that this is strictly below $V^{EAO}$. The grey line shows the highest profit from a single guaranteed delivery offer. In \Cref{C:EAOvsEASI} and its proof, we show that this is strictly below the profit achievable only with at-will offers, and both are strictly below $V^{EAO}$.

\begin{figure}[H]
\centering
\includegraphics[width=0.8\textwidth]{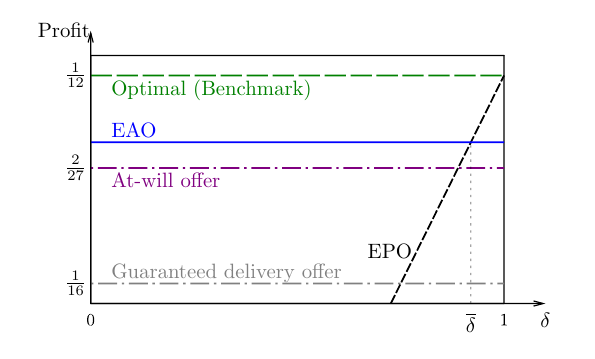}
\caption{Optimal static price mechanisms for uniform distributions.}
\end{figure}

\subsection{Dynamic Price Mechanisms}

\subsubsection{D-mechanisms}

In this section, we consider dynamic price mechanisms. We show that for a sufficiently large discount factor $\delta$, the optimal price mechanism is a dynamic price mechanism. To prove this, it is sufficient to find a particular dynamic price mechanism that gives strictly higher profit than any of the static price mechanisms, meaning strictly higher profit than $V^{static} = \max\{V^{EAO}, V^{EPO}\}$.

Let us define the following simple dynamic mechanism. 
\begin{definition}
D-mechanism with reserve $r$ is a sequence of guaranteed delivery offers: $(p_0, p_1(\omega))$, such that
\begin{enumerate}
\item If the buyer accepts the first offer $p_0$, then trade happens with certainty at time 0 at price $p_0$.
\item If the buyer rejects this offer, the seller makes a new offer $p_1(\omega) \geq r$, which is again a guaranteed delivery offer, but for time 1.
\item If the buyer rejects the second offer as well, then trade does not happen.
\end{enumerate}
\end{definition}

In terms of a reduced-form mechanism, this mechanism has a garbing device with binary signal ($\gB=\{0,1\}$), such that $b \in \{0,1\}$ tells the seller whether the buyer took the first offer or not. For example, if $m(\theta) = \one[\theta \geq \otheta]$ for some threshold $\otheta$, then $q_0(b,\omega)=b, t_0(b,\omega)=b p_0$, $q_1(\theta,\omega) = \one[p_1(\omega) \leq \theta < \otheta]$, and $t_1(\theta,\omega) = p_1(\omega) \one[p_1(\omega) \leq \theta < \otheta]$.

The following proposition shows that such mechanism has its merit. 

\begin{proposition} \label{P:DvsEPO}
  Profit from the D-mechanism with $r=0$ is strictly higher than the profit from the optimal ex-post price mechanism for any $\delta < 1$.
\end{proposition}	
The proof is in \Cref{A:DvsEPO}

Using this proposition, we can immediately conclude for sufficiently high $\delta$, the optimal price mechanism is dynamic.

\begin{corollary} 
  With $\delta \geq \deltastatic$, the maximal profit from a D-mechanism with reserve $r$ is strictly higher than the profit from the optimal static price mechanism.
\end{corollary}	

\begin{proof}
Let $V^D(r)$ be the profit from the D-mechanism with any reserve $r$, and let $r^D = \argmax V^D(r)$. Then clearly the maximal profit from D-mechanism with reserve $V^D(r^D) \geq V^D(0)$. \Cref{P:DvsEPO} shows that $V^D(0) > V^{EPO}$ for all $\delta$, and \Cref{C:optimal_static} shows that $V^{static} = V^{EPO}$ for all $\delta \geq \deltastatic$. Therefore, for all $\delta \geq \deltastatic$,
$V^D(r^D) \geq V^D(0) > V^{EPO} = V^{static}$.
\end{proof}	

The following figure illustrates these results in the case of uniform $F$ and $G$ distributions.\footnote{Calculations for $V^D(r^D)$ are available upon request.}

\begin{figure}[H]
	\centering
	\includegraphics[width=0.8\textwidth]{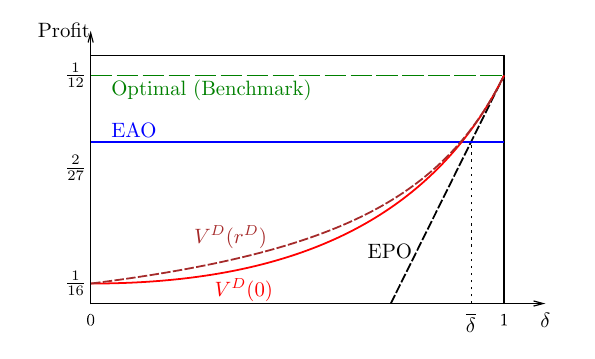}
	\caption{D-mechanisms compared with static price mechanisms for uniform distributions.}
\end{figure}

\subsubsection{Additional Requirements for Price Mechanisms}

We end this section with an example that illustrates that even the requirements we imposed on price mechanism, i.e., mechanisms that are deterministic and satisfy (NPWT) property, are not sufficient to guarantee that the seller cannot approximate a two-part tariff closely if we allow for dynamic price mechanisms.
\paragraph{Example 2.} Consider the following fully revealing price mechanism. Let $q_0(\theta,\omega) = \one[\psi(\theta) \geq \omega]$, i.e., the ex-ante allocation rules are the optimal one, and let $q_1(\theta,\omega) = \one[0 \leq \psi(\theta) < \omega]$, i.e., if the buyer has a positive virtual surplus and was not served at time 0, she will be served at time 1 with certainty. This is certainly a suboptimal allocation, but when $\delta$ is small, the distortion from serving the buyer with certainty at time 1 is small. Now, to make this allocation rule incentive compatible for both parties, we set $t_0(\theta,\omega) = q_0(\theta,\omega) p_0(\theta)$ and $t_1(\theta,\omega) = q_1(\theta,\omega) p_1(\theta)$, where $p_0(\theta)$ and $p_1(\theta)$ are prices determined below. This mechanism is a price mechanism by construction.

We begin with the seller's incentives. Obviously when $\psi(\theta)<0$, there is no trade, so no transfers. Let's fix $\psi(\theta) \geq 0$. Seller's profit when announcing $\omega'$ when the true values are $\theta,\omega$ is now
\[
  v(\omega'|\theta,\omega)
  = 
  \begin{cases}
    p_0(\theta) - \omega
    & \text{ if }\omega' \leq \psi(\theta) \\
    \delta[p_1(\theta)-\omega] 
    & \text{ if }\omega' > \psi(\theta)    
  \end{cases}
\]
The seller type $\omega=\psi(\theta)$ must be indifferent between the two options, which gives us
$p_0(\theta) = \delta p_1(\theta) + (1-\delta) \psi(\theta)$.

Now we see that $p_1(\theta)$ plays a role quite similar to that of  the fixed payment in the two-part tariff. The transfer includes $\delta p_1(\theta)$ no matter what the seller reports, but any report inducing the early trade means additional payment $(1-\delta) \psi(\theta)$. As such report comes with additional cost $(1-\delta) \omega$ (early trade instead of late), it is incentive compatible for the seller to report the type truthfully. 
To determine $p_1(\theta)$ we look at the buyer's incentives. Notice that the expected allocation rule for buyer who reports $\theta'$ such that $\psi(\theta) \geq 0$ is
\begin{align*}    
  Q(\theta') 
  = \int_0^1 [q_0(\theta',\omega) + \delta q_1(\theta',\omega)] dG(\omega)
  = G(\psi(\theta')) + \delta [1-G(\psi(\theta'))]
\end{align*}
Specifically, the expected transfers need to satisfy
\begin{align*}
  T(\theta) 
  &= \theta Q(\theta) - \int_{\theta^*}^{\theta} Q(x) dx.
\end{align*}
On the other hand, we can use $p_0$ and $p_1$ to compute $T$ to get
\[
  T(\theta) 
  = \int_0^{\psi(\theta)} p_0(\theta) dG(\omega) + \int_{\psi(\theta)}^1 \delta p_1(\theta) dG(\omega)
  = \delta p_1(\theta) + (1-\delta) \psi(\theta) G(\psi(\theta)).
\]
Combining the two expressions for $T(\theta)$, we get
\begin{align*}
\delta p_1(\theta)
&=
\theta \left( G(\psi(\theta)) + \delta [1-G(\psi(\theta))] \right) 
- (1-\delta) \psi(\theta) G(\psi(\theta)) 
\\
&- \int_{\theta^*}^{\theta} \left( G(\psi(x)) + \delta [1-G(\psi(x))] \right) dx.
\end{align*}
Finally, note that 
\[
\lim_{\delta \to 0} \delta p_1(\theta)
=
\theta G(\psi(\theta))
- \psi(\theta) G(\psi(\theta)) 
- \int_{\theta^*}^{\theta} G(\psi(x)) dx 
= \lambda(\theta),
\]
where $\lambda(\theta)$ is the fixed component of the two-part tariff in the optimal (general) mechanism. Similarly, $\lim_{\delta \to 0} p_0(\theta) = \psi(\theta)$. 
Therefore, as $\delta \to 0$, the mechanism gets closer and closer to the optimal (general) mechanism. 

This example illustrates two facts. First, the class of price mechanisms that we are studying is rich enough that the seller has many levers to achieve profits close to the benchmark profit, even when we put some natural restrictions on the mechanisms. Second, as in Example 1, such mechanisms are typically not more realistic or simpler than the original two-part tariffs, because they require, when $\delta $ becomes arbitrarily small, unboundedly large transfers by the buyer.

To avoid mechanisms such as these, that rely on the dynamic nature of the problem, but do so in an unrealistic way, it seems natural to restrict the class of mechanisms further. We call these ex-post individually rational price mechanisms. Informally, they require not only that transactions happens deterministically and that the money changes hands only when the good is delivered, but also that at the moment the transaction happens, the buyer must not be unhappy with the transaction.

\begin{definition}
An ex-post individually rational price mechanism is a price mechanism such that $t_0(b,\omega) \leq \theta$ and $t_1(\theta,\omega) \leq \theta$ for all $b,\theta,\omega$. We denote with $\gM^{EPIRP} \subset \gM^P$ the class of ex-post individually rational price mechanisms.
\end{definition}

\section{Extensions} 
\label{S: extensions}
In this section, we consider two natural questions that arise from our setting. Up to this point, we've assumed that the seller learns her type with delay, but there is no cost of learning. We call this the case of \emph{free learning}. In some settings, however, it may make more sense to assume that to learn her type, the seller might have to pay a cost. For example, if we interpret the seller's type as a production cost, then the seller might need to make an investment in research and development to learn what such production cost is. In these cases, the seller cannot provide the good unless that investment is made. We call this the \emph{necessary learning} case. In other situations, the good might be already available to the seller, but she may not be able to attribute it a precise valuation, so that learning the cost requires hiring an expert or expanding some effort to learn her type. In these cases, the seller can still provide the good even if she decides not to spend to learn her valuation. We call this the \emph{optional learning} case.

We study both these costly learning models in \cref{ss:cost_learning}. Our analysis focuses again on optimal general mechanisms and shows, as is intuitive, that if the learning costs are too high, the seller either does not offer a sale (in the case of necessary learning) or offers a price with guaranteed delivery because she chooses not to learn her type (in the case of optional learning). If, however, these costs are low enough, then the optimal mechanisms still feature the two-part tariff property of the case where learning is free. This implies that our results are qualitatively robust to the introduction of learning costs. 

An important difference between the two cases is that in the case of necessary learning (as when learning is free), the seller only offers prices that are contingent on her type realization: whatever price the buyer accepts, trade is not guaranteed because this price needs to be above the seller's type. In the case of optional learning, the seller decides to learn her type only when the buyer type is not too high or too low and offers these buyer types a price contingent on her type realization. However, with optional learning, when the seller learns that the buyer's type is sufficiently high,  she will prefer to offer a price with guaranteed delivery (because her type is likely to be below this price) over learning her cost.  

In subsection \ref{ss: efficiency} we imagine that the mechanism designer's objective is ex-post efficiency rather than maximizing the seller's profit. As is well-known, \cite{myerson_efficient_1983} show that ex-post efficiency cannot be guaranteed when both buyer and seller face Bayesian incentive compatibility constraints and interim individual rationality constraints. In our setting, the natural counterpart is to ask the same question when the buyer faces Bayesian incentive compatibility and interim individual rationality constraints while the seller faces dominant strategy incentive compatibility constraints. In addition, since we are taking the view of the seller as an agent rather than the principal, we impose ex-ante individual rationality for her or, at least, that the seller wishes to participate upon learning the buyer's type. Our result shows that, contrary to the \cite{myerson_efficient_1983}, a mechanism that guarantees ex-post efficiency \emph{does exist}. This suggests that the slackening of the individual rationality constraint for the seller more than compensates for the hardening of her incentive compatibility constraint. Further, the mechanism that implements the ex-post efficient allocation also relies on two-part tariffs, except that now the buyer always receives a subsidy from the seller.\footnote{If, in our setting, we were maximizing the buyer's expected utility, then the benchmark allocation could be implemented \emph{without} a two-part tariff. 
}

\subsection{Costly Learning}
\label{ss:cost_learning}
As discussed above, we distinguish between the case where in order to deliver the good, the seller must pay a cost to learn $\omega$ from the case where the good can be delivered even if the seller does not pay the cost to learn $\omega$. We begin with the former case and again work throughout with fully-revealing mechanisms.\footnote{Clearly the presence of learning costs, does not change the argument for working with these.}
\subsubsection{Necessary learning}
\label{necessary_learning}

We follow the approach used in subsections \ref{SS:freelearning} and \ref{SS:optimality}, of characterizing first the benchmark allocation under the assumption that the seller reveals her type truthfully (if she decided to pay to learn her cost) and then show that this allocation can be also obtained in a setting where the seller cannot commit to tell the truth. 

\paragraph{Benchmark mechanism.} We assume that, in order to learn $\omega$, S must pay a cost $c>0$. This decision is made \emph{after} the seller observes the buyer's valuation. We assume this particular timing because it seems realistic to allow the seller to decide not to learn her cost if she learns that with a particular buyer, a trade won't take place regardless of the seller's type realization. 
The proposition below characterizes the benchmark mechanism:
\begin{proposition}
    \label{P: necessary_learning_benchmark}
    In the benchmark case of necessary learning with a cost $c>0$, the optimal allocation is:
\begin{align*}
  q_0(\theta,\omega) = \one[c \leq \gG(1)] 
  \cdot
  \one[\max(\omega,\gG^{-1}(c)) \leq \psi(\theta)]
\end{align*}
and the respective transfers are
\begin{align*}
  q_0(\theta,\omega) = 
  \one[c \leq \gG(1)] 
  \cdot
  \one[\max(\omega,\gG^{-1}(c)) \leq \psi(\theta)]
  \cdot
  \psi^{-1}((\max(\omega,\gG^{-1}(c))).
\end{align*}

The seller payoff is zero if $c > \gG(1)$, while if $c \leq \gG(1)$ we have
\begin{equation*}
  V^{BNC}(c) 
  = \int_{\psi^{-1}(\gG^{-1}(c))}^1 \left[-c+\int_0^{\psi(\theta)} \left(
    \psi(\theta) - \omega  
  \right) dG(\omega)\right]dF(\theta).
\end{equation*}	
\end{proposition}
The proof is in Appendix \ref{SS: proof_necessary_learning_benchmark}.

The case where $c \geq \gG(1)$ is very intuitive: if the cost of learning is too high for any possible type of buyer, the seller does not expect her gains from trade to cover the cost of learning and so, from an ex-ante perspective, it is optimal to never trade. If $c < \gG(1)$, then some trade is possible, but the menu of prices is different from the benchmark where $c=0$. In particular, she commits to selling at $\psi^{-1}(\omega)$ as long as her type turns out to be no smaller than $\gG^{-1}(c)$. If her type is smaller than $\gG^{-1}(c)$, then the price is fixed at $\psi^{-1}(\gG^{-1}(c))$ because the seller must also take the cost of learning into account.
Since the decision on whether to pay $c$ or not is made after the seller learns the buyer type, the price schedule the seller chooses is the same as the one she chooses when $c=0$, but it is truncated at $\psi^{-1}(\gG^{-1}(c))$. Finally, since both $\psi^{-1}(x)$ and $\gG^{-1}(c)$, are strictly increasing functions with $\psi^{-1}(1)=1$ and $\gG^{-1}(0)=0$, it is easy to see that $V^{BNC}(c)$ is a decreasing function of $c$ with $V^{BNC}(0)=V^B$ and $V^{BNC}(\gG(1))=0$.\footnote{Note that trade occurs if $\theta \geq \max(\gG^{-1}(c),\omega)$ while $c$ is paid if $\theta \geq \gG^{-1}(c).$}

\paragraph{Optimal Mechanism.} If we now move to the case where the seller cannot commit to reveal her type truthfully, the first observation is that if $c > \gG(1)$, then in the benchmark case, it is optimal for the seller not to trade at all, and this clearly remains the case here. If $c \leq \gG(1)$, on the other hand, the analogy with the case where learning is free that we noted in the benchmark case continues, and the optimal mechanism is a two-part tariff, where the set of buyer types the seller wishes to potentially trade with is truncated below, and the $\lambda$ function is adjusted accordingly:

\begin{proposition} \label{P:necessary_learning_optimal}
Suppose $c \leq \gG(1)$. In the case of necessary learning when the seller cannot commit to truthfully revealing her type, there is an optimal mechanism that implements the benchmark allocation: 
	\begin{align*}
        q_0(\theta,\omega) = \one[\max(\omega,\gG^{-1}(c)) \leq \psi(\theta)] 
\end{align*}
with transfers 
\begin{align*}
 t_0(\theta,\omega) = \widehat{\lambda}(\theta) + \psi(\theta) \one [\max(\omega,\gG^{-1}(c)) \leq \psi(\theta)],
	\end{align*}
 where \begin{equation*}
     \widehat{\lambda}(\theta)= (\theta - \psi(\theta))G(\psi(\theta)) - \int_{\psi^{-1}(\gG^{-1}(c))}^\theta G(\psi(y)) dy,
 \end{equation*} whenever $\theta \geq \psi^{-1}(\gG^{-1}(c))$ and $\widehat{\lambda}(\theta)=0$ otherwise.
\end{proposition}
This result is proven following the same steps as in the proof of \cref{P:optimal}. Following such steps one just needs to be clear that $\widehat{\lambda}(\theta) \neq 0$ only applies for $\theta \geq \psi^{-1}(\gG^{-1}(c))$. Finally, that $\int_0^1 \widehat{\lambda}(\theta)dF(\theta)=0$ can be shown using the procedure in the proof of \cref{C:properies_v}. 

The mechanism, as in that setting, asks each buyer type for which trade is possible to pay a price equal to their virtual valuation, but since $\psi^{-1}(\gG^{-1}(c)) > \psi^{-1}(0)$, the set of buyer types that is allowed to trade is limited from those that had a positive virtual valuation to those whose virtual valuation is higher than $\gG^{-1}(c)$. These are the types for which it is worthwhile for the seller to invest $c$ in order to give trade a chance. Also, as in the free learning setting, those buyer types that have a positive chance of trade need to be given the incentive to accept to trade under the conditions of the benchmark allocation and that requires fee/subsidy $\widehat{\lambda}(\theta)$. This latter function still maintains differentiability and the property that:
\begin{equation*}
    \int_0^1 \widehat{\lambda}(\theta)dF(\theta)=\int_{\psi^{-1}(\gG^{-1}(c))}^1 \widehat{\lambda}(\theta)dF(\theta)=0
\end{equation*}
so that, as one would expect, the expected utility for the seller in the optimal mechanism is the benchmark expected utility $V^{BNC}(c)$. It also remains true that $\widehat{\lambda}(1) < 0$ but because of the truncation we have $\widehat{\lambda}(\psi^{-1}(\gG^{-1}(c))) > 0$.\footnote{The proof of these claims about the $\widehat{\lambda}(\theta)$ function is immediate and is available upon request.}. Finally, the immediate indirect implementation is once again a menu of payment pairs, but with a lower bound $\gG^{-1}(c)$ on the prices that the buyer is allowed to offer: $\{(p,\widehat{\Lambda}(p)\}_{p \in [\gG^{-1}(c),1]}$, where $\widehat{\Lambda}(p)=\widehat{\lambda}(\psi^{-1}(p))$. As in the free learning case, for each menu $\widehat{\Lambda}(p)$ represents a fee/subsidy that is paid regardless of trade, while trade occurs at price $p$ if the seller agrees to deliver.

\subsubsection{Optional learning}
\label{optional_learning}
We now look at the case where the seller still pays a cost $c>0$ to learn her type, but where this is not necessary for trade: if she does not pay $c$, the seller can still provide the good to the buyer, but won't know what her ex-post utility is. In this case, the rationale for the seller to learn her type is not to be able to trade but to be able to tailor the allocation rule to both $\theta$ and $\omega$, otherwise, the allocation must solely depend on the former. 

\paragraph{Benchmark mechanism.} We begin again with the benchmark case and provide a counterpart to propositions \ref{P:benchmark} (the free learning case) and \ref{P: necessary_learning_benchmark} (the costly and necessary learning case): 

\begin{proposition}
    \label{P: optional_learning_benchmark}
    In the benchmark case of optional learning with a cost $c>0$, the seller will not pay the cost $c$ whenever $c> \gG(\mathbb{E}(\omega))$. In that case, the optimal allocation is:
    \begin{align*}
    q_0(\theta,\omega) = \one[\mathbb{E}(\omega) \leq \psi(\theta)],
    \end{align*}
    where $\mathbb{E}(\omega)$ is the expected value of $\omega$. This can be implemented by transfers 
    \begin{align*}
        t_0(\theta,\omega) = \psi^{-1}(\mathbb{E}(\omega)) \cdot \one[\mathbb{E}(\omega) \leq \psi(\theta)].
\end{align*}
If $c \leq \gG(\mathbb{E}(\omega))$ then the seller will pay the cost if and only if $\theta \in [\psi^{-1}(\gG^{-1}(c)),\chi^{-1}(c)]$, where 
\begin{equation*}
   \chi(\theta)=\mathbb{E}(\omega)+\gG(\psi(\theta))-\psi(\theta),
\end{equation*}
so that $\psi^{-1}(\gG^{-1}(c)) \geq \psi^{-1}(0)$ and $\chi^{-1}(c) \leq 1$. The optimal allocation is
\begin{align*}
q_0(\theta,\omega) = \one[\omega \leq \psi(\theta)] \cdot \one[\theta \in [\psi^{-1}(\gG^{-1}(c)),\chi^{-1}(c)] + \one[\theta > \chi^{-1}(c)].
\end{align*}
This can be implemented by transfers
\begin{gather*}
t_0(\theta,\omega) =\rho(\theta)\cdot \one[\omega \leq \psi(\theta)] \cdot \one[\theta \in [\psi^{-1}(\gG^{-1}(c)),\chi^{-1}(c)] + \overline{\rho}\cdot \one[\theta > \chi^{-1}(c)],
\end{gather*} where
\begin{gather*}
\rho(\theta)=\left(\theta-\frac{\int_{\psi^{-1}(\gG^{-1}(c))}^{\theta} G(\psi(y))dy}{G(\psi(\theta)}\right) \; \text{and} \;\; \overline{\rho} =\left(\chi^{-1}(c) - \int_{\psi^{-1}(\gG^{-1}(c))}^{\chi^{-1}(c)} G(\psi(y))dy \right).
\end{gather*}
The expected utility for the seller from this mechanism if $c>\gG(\mathbb{E}(\omega))$ is
\begin{gather*}
  V^{BOC}(c) = \int_{\psi^{-1}(\mathbb{E}(\omega))}^1 \int_0^1 
 \left(
    \psi(\theta) - \omega  
  \right)  dG(\omega) dF(\theta)
  \end{gather*}
  while if $c \leq \gG(\mathbb{E}(\omega))$ then 
  \begin{align*}
   V^{BOC}(c) 
   &= \int_{\psi^{-1}(\gG^{-1}(c))}^{\chi^{-1}(c)} \left( -c + 
  \int_0^{\psi(\theta)} (\psi(\theta) - \omega)dG(\omega)\right)dF(\theta) 
\\
  &+ \int_{\chi^{-1}(c)}^1 \int_0^1 (\psi(\theta) - \omega)dG(\omega)dF(\theta).
\end{align*}	
\end{proposition}
To parse this result, it is immediately easy to see that if $c>\gG(\mathbb{E}(\omega))$, then the seller never finds it optimal to learn her type, whatever the buyer's type she faces. Therefore, she will sell the good if the buyer's virtual valuation is above her expected type (or, alternatively, at price $\psi^{-1}(\mathbb{E}(\omega))$. Since the seller does not learn her type, the allocation cannot depend on it, and delivery will be guaranteed. Obviously, this also means that $V^{BOC}(c)$ is positive and constant in $c$ for all $c>\gG(\mathbb{E}(\omega))$.

If $c \leq \gG(\mathbb{E}(\omega))$, on the other hand, things are more complicated. First note that the seller observes $\theta$ as reported by the buyer, and given this buyer type, she can decide i) whether there is an opportunity to trade with such type and ii) whether she should pay $c$ to learn her $\omega$ in order to make trade contingent on her type being smaller than the buyer's or whether she should not pay $c$ but commit to trade for sure with this buyer. The function $\gG(\psi(\theta))$ represents the expected utility that the seller can expect from trading with type $\theta$ if she pays $c$ and learns her type. Hence, if $\theta \geq \psi^{-1}(\gG^{-1}(c))$, the seller will prefer to pay $c$, learn her own type and have a positive probability of trading with $\theta$ over not trading with $\theta$ and not learning her type. The function $\chi(\theta)$, instead, represents the benefit that the seller has in learning $\omega$ and then trade with $\theta$ if $\omega \leq \theta$ over not learning $\omega$ and trading with $\theta$ for sure. Thus, if $\theta \leq \chi^{-1}(c)$ then the seller prefers to pay $c$ and leave herself the opportunity to not trade with $\theta$ if it turns out that $\theta < \omega$. This means that, having observed the buyer's type, the seller will not invest in learning her cost if the buyer's type is too low ($\theta < \psi^{-1}(\gG^{-1}(c))$ or too high ($\theta > \chi^{-1}(c)$). In the first case, this is because the seller does not wish to trade with such buyer type. In the second case, instead, the buyer type is so high that the seller prefers to sell with guaranteed delivery because the price she gets is sufficiently high that the possibility that her type turns out to be above such price (where she would benefit from not delivering) is outweighed by the cost of learning the type itself.\footnote{In the proof we show, in particular, that $\chi^{-1}(c) \geq \psi^{-1}(\mathbb{E}(\omega))$ so that whenever $\theta > \chi^{-1}(c)$, automatically $\theta > \psi^{-1}(\mathbb{E}(\omega))$ which is the condition that guarantees that the seller wishes to trade with guaranteed delivery with this buyer type.} 

For intermediate values of the buyer's type, the seller will decide to learn her cost. For such buyer types, the price that she obtains is not too high. Thus, she prefers to pay the cost of learning her type because this allows her to cancel the trade---if unprofitable---over not paying the cost and having to guarantee delivery. In the proof we show that  $\psi^{-1}(\gG^{-1}(c))$ is a strictly increasing function of $c$ with $\psi^{-1}(\gG^{-1}(0))=\psi^{-1}(0)$ and $\psi^{-1}(\gG^{-1}(\gG(\mathbb{E}(\omega)))=\psi^{-1}(\mathbb{E}(\omega))$ while $\chi^{-1}(c)$ is a strictly decreasing function of $c$ with $\chi^{-1}(0)=1$ and $\chi^{-1}(\gG(\mathbb{E}(\omega))=\psi^{-1}(\mathbb{E}(\omega))$ so that the interval $[\psi^{-1}(\gG^{-1}(c)),\chi^{-1}(c)]$ becomes the interval $[\psi^{-1}(0),1]$ when $c=0$. This is intuitive: if the cost of learning her type approaches zero, the seller always learns whenever a buyer's type is one where she may wish to trade, and so the optimal allocation converges to that under free learning. Conversely, the interval disappears when $c=\gG(\mathbb{E}(\omega))$ because the cost of learning her type is higher than any possible benefit she might get, and she will either not trade or trade with guaranteed delivery.

As for the transfers, note that $\rho(\theta)$ in an increasing function with $\rho(\chi^{-1}(c)) < \overline{\rho}$. This is important because $\rho(\theta)$ is a price that the buyer agrees to pay, but trade will only occur if the seller's type is smaller than $\rho(\theta)$, whereas $\overline{\rho}$ is a price that guarantees delivery. So, to be incentive compatible for buyer of type $\chi^{-1}(c)$ to be indifferent between $\rho(\chi^{-1}(c))$ and $\overline{\rho}$, the latter must be sufficiently higher than the former.

Finally, it is easy to see that $V^{BOC}(c) > V^{BNC}(c)$ for any $c>0$. This is quite obvious as the under necessary learning the seller has two options for any possible buyer's type: give him a positive probability of trading and pay $c$ or not give him any chance of trade. In optional learning, the seller always also has the possibility of trading with a particular buyer type and not pay $c$. 

\paragraph{Optimal Mechanism.} As in the case with necessary learning, if we consider the case where the seller cannot commit to revealing her type truthfully under optional learning, the analogy with the case where learning is free is evident. Note first, however, that if $c > \gG(\mathbb{E}(\omega))$, then the benchmark allocation is \emph{independent} of the seller's type, simply because this is unknown. Hence, whether the seller can commit to reveal something she doesn't know is irrelevant and the optimal mechanism is the benchmark mechanism. Thus, the issue is only relevant when $c \leq \gG(\mathbb{E}(\omega))$ so that there is an interval $[\psi^{-1}(\gG^{-1}(c)),\chi^{-1}(c)]$ of buyer types for which the allocation depends on $\omega$. 

\begin{proposition}
    \label{P: optional_learning_optimal}
    Suppose $c \leq \gG(\mathbb{E}(\omega))$. In the case of optional learning, when the seller cannot commit to truthfully revealing her type, there is an optimal mechanism that implements the benchmark allocation:
\begin{align*}
q_0(\theta,\omega) = \one[\omega \leq \psi(\theta)] \cdot \one[\theta \in [\psi^{-1}(\gG^{-1}(c)),\chi^{-1}(c)] + \one[\theta > \chi^{-1}(c)].
\end{align*}
Transfers for this mechanism are:
\begin{gather*}
t_0(\theta,\omega) 
=\psi(\theta) \cdot 
\one[\omega \leq \psi(\theta)]+
\overline{\lambda}(\theta)]
\cdot \one\left[ \theta \in [\psi^{-1}(\gG^{-1}(c)),\chi^{-1}(c)] \right]
+ \overline{\rho} \cdot 
\one[\theta > \chi^{-1}(c)],
\end{gather*} 
where $\overline{\rho}$ is defined in proposition \ref{P: optional_learning_benchmark} while
\begin{gather*}
\overline{\lambda}(\theta)= (\theta -\psi(\theta))G(\psi(\theta) - \int_{\psi^{-1}(\gG^{-1}(c))}^{\theta} G(\psi(y))dy 
\end{gather*}
for all $\theta \in [\psi^{-1}(\gG^{-1}(c)),\chi^{-1}(c)]$ and $\overline{\lambda}(\theta)=0$ otherwise.
\end{proposition}
The difference between the benchmark and the optimal mechanisms is only on the transfers over the interval $\left[\psi^{-1}(\gG^{-1}(c)),\chi^{-1}(c) \right]$. That these transfers are dominant strategy incentive compatible for the seller follows immediately using the same argument as in \cref{P:optimal}. The only other task is to verify that the transfers also equal expected transfers for the buyer, which again is the same as in \cref{P:optimal}.

The benchmark mechanism and the optimal mechanism are the same except that when $\theta \in [\psi^{-1}(\gG^{-1}(c)),\chi^{-1}(c)]$, the price is $\psi(\theta)$ to make sure the benchmark allocation is incentive compatible for the seller. Correspondingly, there is the additional fee or subsidy $\overline{\lambda}(\theta)$ that is paid regardless of trade that serves the purpose of guaranteeing incentive compatibility for buyer types in this region. There is, therefore, at least for these buyer types, a two-part tariff. Note that the truncation of the fee/subsidy function here is from two sides: if $\theta < \psi^{-1}(\gG^{-1}(c))$ then there is no possibility of trade, and no payments are made, whereas if $\theta > \chi^{-1}(c)$, the seller chooses a price that is incentive compatible for the buyer without needing to make sure it is incentive compatible for her future self. Thus, a two-part tariff is no longer needed. One of the consequences is that while with both free learning and necessary learning (when the cost was not excessive), the very highest types always were offered an allocation contingent on the seller's type and always received a subsidy, here the highest seller types are offered a price with guaranteed delivery. This means that since $\chi^{-1}(c)$ is decreasing in $c$, eventually it will become sufficiently small that $\overline{\lambda}(\theta)$ will be positive for all $\theta \in [\psi^{-1}(\gG^{-1}(c)),\chi^{-1}(c)]$. In other words, if $c$ is small but not too small, all types for which the seller decides to pay $c$ and learn her type, will have to pay a positive fee in the optimal mechanism.

Finally, a natural indirect implementation of this mechanism is the following. When $c>\gG(\mathbb{E}(\omega))$ then, as already discussed in the benchmark case, the seller offers the buyer a price $\psi^{-1}(\mathbb{E}(\omega))$ with guaranteed delivery. When $c \leq \gG(\mathbb{E}(\omega))$, the seller offers the buyer \emph{both} the price $\overline{\rho}$ with guaranteed delivery \emph{and} a menu of price-fee or subsidy payments with a lower bound $\gG^{-1}(c)$ \emph{and} an upper bound $\psi(\chi^{-1}(c))$ on the prices that the buyer is allowed to choose: $\{(p,\overline{\Lambda}(p)\}_{p \in [\gG^{-1}(c),\psi(\chi^{-1}(c))]}$, where $\overline{\Lambda}(p)=\overline{\lambda}(\psi^{-1}(p))$. As for the previous mechanisms, $\overline{\Lambda}(p)$ is paid regardless of trade, while the prices do not guarantee delivery: they are contingent on the seller's willingness to deliver.\footnote{Because of this, $\psi(\chi^{-1}(c))<\overline{\rho}$.} 

This implementation is similar to the optimal ex-ante price mechanism in the sense that it offers a menu that includes guaranteed delivery options as well as at-will options. The difference is that in the case of price mechanisms, the at-will elements are distorted due to incentive constraints (due to the inability to use two-part tariffs), whereas in the case of costly learning, the distortion comes from learning costs.

\subsection{Efficiency}
\label{ss: efficiency}

Our focus in the paper is profit maximization for the seller when she learns her information with delay. Our analysis shows that, in general, a mechanism exists that implements the same allocation as in the benchmark cases where either the seller can commit to revealing her information truthfully (eliminating incentive compatibility issues on the seller's side) or when she can commit to ignoring the information revealed by the buyer (equivalent to facing interim incentive compatibility constraints for her future self). However, this mechanism requires payments even when the trade does not occur.

Therefore, it is natural to ask what results an analogous analysis would produce when the objective is ex-post efficiency. This is particularly interesting because in this case, as shown by \cite{myerson_efficient_1983}, we cannot have ex-post efficiency when both buyer and seller face Bayesian incentive compatibility and interim individual rationality constraints, while it can be implemented if those constraints are applied only to one side. 
In this subsection, as we did in section \ref{SS:optimality}, we modify the constraints the seller faces by imposing dominant strategy incentive compatibility. We will also not impose interim individual rationality since this is meaningless when the seller learns with delay. Instead, it makes sense to ask whether she is willing to participate at the ex-ante stage or, more stringently, at a ``reverse" interim stage, where she has learned the buyer's type but not yet her own. As we show below, contrary to the \cite{myerson_efficient_1983} setting, ex-post efficiency is implementable by a two-part tariff mechanism, regardless of which of the two individual rationality constraints for the seller discussed above we impose:
\begin{proposition}
    \label{P: efficiency}
    Under the incentive and individual rationality constraints posed by our setting, the ex-post efficient allocation $q_0(\theta,\omega)=\one[\omega \leq \theta]$ is implementable by a mechanism where transfers are:
\begin{gather*}
t_0(\theta,\omega) =\theta \one[\omega \leq \theta]-\gG(\theta).
\end{gather*}
\end{proposition}
The proof is in \cref{SS: proof_efficiency}. An immediate observation one can make following \cref{P: efficiency} is that while the delay with which the seller learns her type makes us impose a stronger constraint on her incentive compatibility than in \cite{myerson_efficient_1983}, the same structure allows us to weaken the seller's individual rationality in a way that more than compensates and allows for the implementation of ex-post efficiency. Not surprisingly, given this observation, the mechanism requires the seller to provide each buyer type with a subsidy to compensate the latter for choosing a price equal to his type. Thus, all buyer types get a surplus equal to $\gG(\theta)$ and in \cref{SS: proof_efficiency}, we show that the ``reverse" interim individual rationality is tight: upon learning the buyer's valuation, the seller's expected utility is zero.

This result also shows how the seller's constraint in dominant strategies still forces the mechanism to pinpoint exactly what transfers are needed. This means that two-part tariffs are needed, although the fixed component is always a subsidy in this case. 

Obviously, we found this mechanism because the informational environment is very different from that of \cite{myerson_efficient_1983}: if the seller knew her type before deciding whether to participate in the mechanism, then high $\omega$ types would not want to participate. Type $\omega=1$, for example, would expect no gains from trade whatever $\theta$ is but would still have to pay the subsidy. The fact that the seller has to agree \emph{before} she learns her type is crucial for implementing this mechanism.

Finally, the mechanism itself suggests an obvious indirect implementation: the buyer can pick any price $p \in [0,1]$. For any such price, the buyer gets a subsidy $\gG(p)$ from the seller and receives the good at price $p$ if the seller, upon learning her cost, agrees to deliver at that price.  

\section{Discussion} \label{S:discussion}

We studied bilateral trade in situations where the seller wishes to maximize her profit but learns her type with delay. The mechanism must therefore account for the seller's own incentive to take advantage of the information about the buyer she learns. We identify a potential trade-off between contracting earlier with limited information, which creates stronger incentive problems, and postponing the contract to avoid those issues, though at the cost of delay.

Our first main result shows that an optimal mechanism can effectively address these challenges. The optimal mechanism involves a menu of two-part tariffs. Traditionally, two-part tariffs are used in standard contracting situations to incentivize the agent to choose the desired action, with the fixed component capturing the surplus ex-ante. For instance, in a typical textbook model involving an agent with a known downward-sloping demand for a good, a monopolist can capture the full consumer surplus by charging the marginal cost as a marginal price and a fixed fee equal to the induced consumer surplus. In our model, however, the value of two-part tariffs lies in their ability to ensure that the seller's own future incentives are correctly aligned through delivery fees, while the fixed component controls the buyer's incentives to select the appropriate menu item. Consequently, our two-part tariffs exhibit a unique feature: unlike the classic literature, the fixed fee must sometimes be positive (for intermediate buyer types) and sometimes negative (for high buyer types). This is driven by the need to adjust informational rents: intermediate types are offered a relatively low delivery price, resulting in high ex-post information rents. To reduce these information rents, an additional charge is necessary. Conversely, high types are charged a high delivery price, so providing them with the necessary information rents requires an ex-ante subsidy. Our extensions to costly learning demonstrate that this result is robust to changes in the model.

Our second main result focuses on price mechanisms. This analysis is motivated by the observation that the menus of two-part tariffs we identify as optimal are not very common in practice, possibly because the implementation of the two-part tariffs may be too complex. Therefore, we study a more restricted class of mechanisms to find more realistic options. We first characterize the optimal static price mechanisms, where the contract occurs either before or after the cost information is available, without any dynamic screening. If the discount factor is sufficiently high, the optimal static mechanism waits until the cost information is known and then makes a cost-dependent take-it-or-leave-it offer. However, if the discount factor is low, the optimal ex-ante price mechanism involves a menu of offers, including one or more at-will contracts and one guaranteed-delivery offer. The highest buyer types opt for guaranteed delivery at a higher price, while at-will contracts enable the seller to further screen intermediate-type buyers. Our second main result shows that the optimal price mechanism becomes dynamic if the discount factor is large enough. This dynamic mechanism balances trade-offs by offering early delivery to high types, thus reducing delay costs, and late delivery to low types, thus reducing incentive problems.

We extend our analysis of general mechanisms in two ways: we first consider the possibility that learning her type is costly for the buyer. We distinguish the case where the seller needs to learn her type in order to trade from the case where she can still trade even if she does not learn.  Our results show that the analysis for the case where learning is free, is essentially robust in that when learning the cost is necessary, the seller either decides not to trade at all (if the cost is too high), or uses a two-part tariff. When learning is optional, the seller either chooses to trade at a fixed price without learning the cost (if the cost is too high) or chooses to trade in the same way with high buyer types, but learns the cost and uses a two-part tariff with intermediate buyer types.

Finally, we also look at the issue of implementation of ex-post efficiency. Our buyer faces the same constraints as in the classic \cite{myerson_efficient_1983} setting, while the seller faces stronger incentive and weaker participation constraints. We show that, again using a two-part tariff, \emph{there is} a mechanism that implements ex-post efficiency. In this case, however, the fixed component is always a subsidy: here the mechanism is such that the seller is always forced to give up all the surplus to the buyer.

A key limitation of our analysis is that while we demonstrated that dynamic price mechanisms can sometimes be optimal, we did not explicitly characterize what these optimal mechanisms look like. Identifying them remains an open question that requires further exploration. More broadly, this work represents an initial step towards a larger agenda in mechanism and information design, focusing on models where the principal actively participates in the designed game because she will have information, but with delay. While similar concepts, such as credible auctions and renegotiation-proof games, have been explored in the literature, there is still substantial work to be done.

\bibliographystyle{apecon}
\bibliography{toomash-pip}

\appendix
\section{Proofs} \label{S:proofs}
\subsection{Proofs for \Cref{S:model}} 
\subsubsection{Proof of \Cref{T:revelationprinciple}} \label{A:revelationprinciple}

First, let us introduce the notation for mixed strategies.
\begin{itemize}
\item $\beta_0(\theta)$ is a random variable, with distribution $\Beta_0(b_0|\theta)$, $\supp \beta_0(\theta) \subset \gB_0$ is its support and $\expect_{\beta_0(\theta)}[\cdot]$ is the expectation over $\gB_0$ with respect to $\beta_0$ for a given $\theta$.\footnote{
With continuous distributions, $\expect_{\beta_0(\theta)}[f(b_0)]
= \int_{\gB_0} f(b_0) d\Beta_0(b_0|\theta)$, and an equivalent of this with mass points.
}
\item Similarly, $\supp \sigma(\omega,b_0)$, $\Sigma(s|\omega,b_0)$, $\expect_{\sigma(\omega,b_0)}[\cdot]$, for each $\omega$ and $b_0 \in \gB_0$.
\item Similarly, $\supp \beta_1(\theta,b_0,s)$, $\Beta_1(b_1|\theta,b_0,s)$, $\expect_{\beta_1(\theta,b_0,s)}[\cdot]$, for each $\theta, b_0,s$.
\end{itemize}
Fix a mechanism and an equilibrium:
\begin{itemize}
\item $\eqBi(\theta|b_0,s) = \expect_{\beta_1(\theta,b_0,s)} q(b_0,s,b_1)$ is the ex-post expected net present value of the quantity, conditional on history $(b_0,s)$ and buyer type $\theta$ and $\etBi(\theta|b_0,s)$ is defined analogously.
\item $\eqS(\theta,\omega|b_0) = \expect_{\sigma(\omega,b_0)} \eqBi(\theta|b_0,s)$ and $\etS(\theta,\omega|b_0)$ is defined analogously.
\item $\eqBo(\theta,\omega) = \expect_{\beta_0(\theta)} \eqS(\theta,\omega|b_0)$ and an analogous $\etBo(\theta,\omega)$.
\end{itemize}
In other words, these are allocation and transfer functions (NPV of each), conditional on choices made before each decision node, fixing type profiles $\theta,\omega$, and assuming equilibrium behavior starting from period $t$.
These objects depend on true values $(\theta,\omega)$, but can be defined in the same way for situations where some type mimics the behavior of another type. In other words, for a fixed mechanism and an equilibrium, these objects are defined for B, who behaves ``as if'' he is type $\theta$, and S, who behaves as if being ``$\omega$''.

Next, we introduce the notation for the equilibrium payoffs at each history, assuming equilibrium behavior starting from this point onwards:
\begin{itemize}
\item 
$\euBi(\theta|b_0,s) 
= \theta \eqBi(\theta|b_0,s)-\etBi(\theta|b_0,s)
$ is the utility after history $(b_0,s)$ for the buyer of type $\theta$.
\item $\evS(\omega|b_0) 
= \int_{\Theta} [\etS(\theta,\omega|b_0)-\omega \eqS(\theta,\omega|b_0)] dF(\theta|b_0)
$ is the profit after history $b_0$ for the seller of type $\omega$.
\item $\euBo(\theta) 
= \int_{\Omega} [\theta q^0(\theta,\omega)-t^0(\theta,\omega)] dG(\omega)
$ is the utility of the buyer of type $\theta$ before any choices.
\end{itemize}

With this notation, we can compactly define the equilibrium in mixed strategies\footnote{These conditions are direct generalizations of the corresponding conditions for pure strategies in the main text, we use the same labels.}
$(\Beta_0,\Sigma,\Beta_1)$ is an equilibrium in (mixed) strategies, if
\begin{align}
  \euBo(\theta) &\geq 0, \forall \theta, \tag{IRB\textsubscript{0}} \label{E:IRBoMS}
  \\ 
  \euBo(\theta) &\geq \int_{\Omega} [\theta \eqS(\theta,\omega|b_0) - \etS(\theta,\omega|b_0)] dG(\omega), \forall \theta,
  \tag{ICB\textsubscript{0}} \label{E:ICBoMS}
	\\ 
  \evS(\omega|b_0) &\geq 	
  \int_{\Theta} [\etBi(\theta|b_0,s)-\omega \eqBi(\theta|b_0,s)] dF(\theta|b_0),
\forall b_0, \omega,
  \tag{ICS\textsubscript{0}} \label{E:ICSoMS}
	\\ 
  \euBi(\theta|b_0,s)&\geq
  \theta q(b_0,s,b_1) - t(b_0,s,b_1),
\forall b_0, s, \omega,  
  \tag{ICB\textsubscript{1}} \label{E:ICBiMS}
\end{align}
where \eqref{E:ICBoMS}, \eqref{E:ICSoMS}, and \eqref{E:ICBiMS} hold as equalities whenever the corresponding value is on the support. For example, \eqref{E:ICBoMS} is an equality for all $b_0 \in \supp \beta_0(\theta)$.

\begin{proof}[Proof of \Cref{T:revelationprinciple}]
We prove the result in three steps. The first, lemma \ref{L:lemma1B}, shows that the buyer's last message space $\gB_1$ can be without loss of generality replaced by type space $\Theta$. The second, lemma \ref{L:lemma1S} shows similarly that the seller's message space $\gS$ can be replaced by type space $\Omega$. Finally, lemma \ref{L:lemma1Bo} constructs a reduced-form mechanism corresponding to a general mechanism using a garbling device.

\begin{taggedlemma}{1.B1} \label{L:lemma1B}
For any mechanism $\gM$ and an equilibrium $(\Beta_0,\Sigma,\Beta_1)$, there exists an alternative mechanism $\hgM$ and a corresponding equilibrium $(\hBeta_0,\hSigma,\hBeta_1)$, that is payoff-equivalent to the original equilibrium in $\gM$, and
\begin{enumerate}
\item $\hgB_0=\gB_0, \hgS = \gS$, $\hgB_1 = \Theta$,
\item $\hBeta_0=\Beta_0, \hSigma=\Sigma$, and $\beta_1(\theta,b_0,s) = \theta$.
\end{enumerate}
\end{taggedlemma}
	
	\begin{proof}
	Take any mechanism $\gM = (\gB_0,\gS,\gB_1,q_0,t_0,q_1,t_1)$ and an equilibrium 
	$(\Beta_0,\Sigma,\Beta_1)$. 
	We construct a new mechanism: $\hgM$ such that it mimics the same mechanism but with type announcement in the last step. That is, the message spaces are $\hgB_0=\gB_0$, $\hgS = \gS$, and $\hgB_1=\Theta$, and 
	$\hq_0(b_0,s) = q_0(b_0,s)$, 
	$\htr_0(b_0,s) = t_0(b_0,s)$, 
	$\hq_1(b_0,s,\theta) = \expect_{\beta_1(\theta,b_0,s)}q_1(b_0,s,b_1)$,
	and
	$\htr_1(b_0,s,\theta) = \expect_{\beta_1(\theta,b_0,s)} t_1(b_0,s,b_1)$.
	
	We claim that the following strategy profile $(\hBeta_0,\hSigma,\hBeta_1)$ is an equilibrium in $\widehat{\gM}$:
	\[
	\hBeta_0(\theta) = \Beta_0(\theta),\;\;\;
	\hSigma(\omega,b_0) = \Sigma(\omega,b_0),\;\;\;
	\hbeta_1(\theta,b_0,s) = \theta
	\]
	and it gives the same expected payoffs for both players.
	
	Using the same notation as above, but adding a hat for the objects associated to mechanism $\hgM$ and denoting the type declaration by $\theta'$, we get that
	\begin{align*}
	\hq(b_0,s,\theta')
	&= \hq_0(b_0,s)	+ \delta \hq_1(b_0,s,b_1) 
	= \expect_{\beta_1(\theta',b_0,s)} q(b_0,s,b_1),
	\\
	\heqBi(\theta|b_0,s) 
	&= \expect_{\hbeta_1(\theta,b_0,s)} \hq(b_0,s,\theta')	
	= \hq(b_0,s,\theta)	
	= \expect_{\beta_1(\theta,b_0,s)} q(b_0,s,b_1)
	= \eqBi(\theta|b_0,s),
	\\
	\heqS(\theta,\omega|b_0)
	&= \expect_{\hsigma(\omega,b_0)} \heqBi(\theta|b_0,s)
	= \expect_{\sigma(\omega,b_0)} \eqBi(\theta|b_0,s) 
	= \eqS(\theta,\omega|b_0),
	\\
	\heqBo(\theta,\omega) 
	&= \expect_{\hbeta_0(\theta)} \heqS(\theta,\omega|b_0)
	= \expect_{\beta_0(\theta)} \eqS(\theta,\omega|b_0)
	= \eqBo(\theta,\omega).
	\end{align*}
	Transfers have the same property. Therefore also the payoffs $\heuBi=\euBi,\hevS=\evS$, and $\heuBo=\euBo$. This establishes the 
	payoff-equivalence. Finally, we need to verify that the new strategy profile satisfies the equilibrium conditions \eqref{E:IRBoMS}, \eqref{E:ICBoMS}, \eqref{E:ICSoMS}, and \eqref{E:ICBiMS}. The first three are immediate, as they only depend on objects  that are unchanged. For \eqref{E:ICBiMS} 
	\begin{align*}	
	\heuBi(\theta|b_0,s) 
	&\geq
	\theta \hq(b_0,s,\theta') - \htr(b_0,s,\theta')
	,\text{ with ``='' for }\theta' = \theta
	\iff \\
	\euBi(\theta|b_0,s) 
	&\geq 
	\expect_{\beta_1(\theta',b_0,s)} [\theta q(b_0,s,b_1) - t(b_0,s,b_1)]
	,\text{ with ``='' for }\theta' = \theta.
	\end{align*}
	As $\gM$ satisfies \eqref{E:ICBiMS} for all $b_0$, we have that for all $b_1 \in 
	\gB_1$,
	\begin{equation} \label{E:ICBieq}
	\euBi(\theta|b_0,s) 
	\geq 
	\theta q(b_0,s,b_1) - t(b_0,s,b_1),
	\end{equation}
	so the inequality must also hold in expectation. Moreover, for each $b_1 \in 
	\supp \beta_1(\theta,b_0,s)$, \Cref{E:ICBieq} holds as an equality, therefore 
	taking an expectation over $\beta_1(\theta,b_0,s)$ gives also an equality. This 
	establishes that \eqref{E:ICBiMS} is still satisfied. 
	\end{proof}
	
	\begin{taggedlemma}{1.S} \label{L:lemma1S}
	For any mechanism $\gM$ and an equilibrium $(\Beta_0,\Sigma,\Beta_1)$, such that $\gB_1=\Theta$ and $\beta_1(\theta,b_0,s)=\theta$, there exists an alternative mechanism $\hgM$ and a corresponding equilibrium $(\hBeta_0,\hSigma,\hBeta_1)$, that is payoff-equivalent to the original equilibrium in $\gM$, and
	\begin{enumerate}
	\item $\hgB_0=\gB_0, \hgS = \Omega$, $\hgB_1 = \Theta$,
	\item $\hBeta_0=\Beta_0, \hsigma(\omega,b_0)=\omega$, and $\beta_1(\theta,b_0,s) = \theta$.
	\end{enumerate}
	\end{taggedlemma}
	\begin{proof}
	Take a mechanism $\gM$ and an equilibrium $(\Beta_0,\Sigma,\Beta_1)$ satisfying the conditions in the lemma. We construct a new mechanism: $\widehat{\gM}$ such that it mimics the same mechanism, but with type announcement by S. The message spaces are $\hgB_0=\gB_0$, $\hgB_1 = \gB_1 = \Theta$, and $\hgS=\Omega$, and 
	$\hq_0(b_0,s) = \expect_{\sigma(\omega,b_0)} q_0(b_0,s)$, 
	$\htr_0(b_0,s) = \expect_{\sigma(\omega,b_0)} t_0(b_0,s)$, 
	$\hq_1(b_0,s,\theta) = \expect_{\sigma(\omega,b_0)} q_1(b_0,s,b_1)$,
	and
	$\htr_1(b_0,s,\theta) = \expect_{\sigma(\omega,b_0)} t_1(b_0,s,\theta)$.
		
	We claim that the following strategy profile $(\hBeta_0,\hSigma,\hBeta_1)$ is an equilibrium in $\widehat{\gM}$:
	\[
	\hBeta_0(\theta) = \Beta_0(\theta),\;\;\;
	\hsigma(\omega,b_0) = \omega ,\;\;\;
	\beta_1(\theta,b_0,s) = \beta_1(\theta,b_0,s) = \theta
	\]
	and it gives the same expected payoffs for both players.

	Using the same notation as above, and again using $\omega'$ and $\theta'$ as type announcements, we now have
	\begin{align*}
		\hq(b_0,\omega',\theta')
		&= \hq_0(b_0,\omega') + \delta q_1(b_0,\omega',\theta') 
		= \expect_{\sigma(\omega',b_0)} q(b_0,s,\theta'),
		\\
		\heqBi(\theta|b_0,\omega') 
		&= \expect_{\hbeta_1(\theta,b_0,\omega')} \hq(b_0,\omega',b_1)	
		= \hq(b_0,\omega',\theta)
		= \expect_{\sigma(\omega',b_0)} q(b_0,s,\theta)
		= \expect_{\sigma(\omega',b_0)} \eqBi(\theta|b_0,s),
		\\
		\heqS(\theta,\omega|b_0)
		&= \expect_{\hsigma(\omega,b_0)} \heqBi(\theta|b_0,\omega')
		= \heqBi(\theta|b_0,\omega)
		= \expect_{\sigma(\omega',b_0)} \eqBi(\theta|b_0,s)
		= \eqS(\theta,\omega|b_0),
		\\
		\heqBo(\theta,\omega) 
		&= \expect_{\hbeta_0(\theta)} \heqS1(b_0,\theta,\omega)
		= \expect_{\beta_0(\theta)} \eqS(\theta,\omega|b_0)
		= \eqBo(\theta,\omega).
	\end{align*}
	Again, similar properties hold for transfers. Therefore, payoffs satisfy $\hevS=\evS$ and $\heuBo=\euBo$, which immediately establish the payoff equivalence. Also, \eqref{E:IRBoMS} and \eqref{E:ICBoMS} hold. 
	As $\hq$ and $\heqBi$ now differ, we still have to verify \eqref{E:ICSoMS} and \eqref{E:ICBiMS} in the new mechanism.
	
	Let us consider \eqref{E:ICSoMS} first.
	\begin{align*}
	  \hevS(\omega|b_0)
	  &\geq 
	  \int_{\Theta}
		\left[ 
		  \hetBi(\theta|b_0,\omega') - \omega \heqBi(\theta|b_0,\omega')
		\right]
	  dF(\theta|b_0)
	  , \text{ with ``='' for }
	  \omega' = \omega
	  \iff 
	  \\
	  \hevS(\omega|b_0)
	  &\geq 
	  \expect_{\sigma(\omega',b_0)}
	  \int_{\Theta}
	  \left[ 
	   \etS(\theta|b_0,s) - \omega \eqS(\theta|b_0,s)
	  \right]
	  dF(\theta|b_0)
	  , \text{ with ``='' for }\omega' = \omega.
	\end{align*}
	Note that by the \eqref{E:ICSoMS} in the original mechanism and equilibrium, this inequality holds for each $s$ separately, therefore it also holds in expectation. Moreover, the inequality holds as equality for each $s \in \supp \sigma(\omega,b_0)$, so when $\omega'=\omega$, the condition indeed holds as equality. This establishes the \eqref{E:ICSoMS} condition.
	
	Finally, consider \eqref{E:ICBiMS}.
	\begin{align*}
	\heuBi(\theta|b_0,\omega') 
	&\geq
	\theta \hq(b_0,\omega',\theta')-\htr(b_0,\omega',\theta'),
	\text{ with ``='' for }\theta'=\theta
	\iff 
	\\
	\expect_{\sigma(\omega',b_0)}
	\euBi(\theta|b_0,s) 
	&\geq
	\expect_{\sigma(\omega',b_0)} \left[
	\theta 
	q(b_0,s,\theta')-t(b_0,s,\theta')
	\right],
	\text{ with ``='' for }\theta'=\theta.
	\end{align*}
	Again, by \eqref{E:ICBiMS}, this equation holds for each $s$ separately, so it also holds in expectation. And again, it holds as an equality when $\theta'=\theta$ for each $s$ separately, therefore it also holds as an equality in expectation.
	\end{proof}
	
	We have shown that we can set $\gB_1 = \Theta$ and $\gS = \Omega$.	Setting $\gB_0 = \Theta$ would not be without loss, because then the mechanism would reveal information about the buyer's type to the seller, i.e., if $\beta_0(\theta')=\beta_0(\theta)=b_0$, then $b_0$ does not distinguish $\theta'$ and $\theta$, but reporting the type would distinguish. Therefore the \eqref{E:ICSoMS} constraint would include an expectation over a different distribution. This is the reason we need to introduce a garbling device to the mechanism so that the seller only learns a realized $b_0$ and nothing else from the first message.
	
	\begin{taggedlemma}{1.B0} \label{L:lemma1Bo}
		For any mechanism $\gM$ and an equilibrium $(\Beta_0,\Sigma,\Beta_1)$, such that $\gB_1=\Theta, \gS=\Omega, \beta_1(\theta,b_0,s)=\theta$, and $\sigma(\omega,b_0)=\omega$, there exists a reduced-form mechanism $\gM^R$ and truthtelling equilibrium, that is payoff-equivalent to the original equilibrium in $\gM$.
	\end{taggedlemma}
	\begin{proof}
	  We construct the mechanism $\gM^R = (\gB,\mu,q_0^D,t_0^D,q_1^D,t_1^D)$ as follows:
	  $\gB = \gB_0$, $\mu(b_0|\theta') = \Beta_0(b_0|\theta')$, and 
	  $q_i^D \equiv q_i, t_i^D \equiv t_i$.

	  Clearly if both the buyer and the seller reveal their types truthfully, the mechanism $\gM^R$ ensures the same on-path payoffs as $\gM$ in the original equilibrium. We therefore to verify that truthtelling is an equilibrium.
	  
	  As the mechanism starting from a particular realization of $b_0$ is unchanged, revealing $\omega$ is still optimal for the seller. Moreover, by construction, the on-path payoff for B is unchanged, so \eqref{E:IRBoMS} is also satisfied. 
	  
	  We, therefore, only need to ensure that the does not have profitable deviations in terms of $b_0$. For this, note that with $\gM$, B has three types of potential deviations when choosing $b_0$.
	  \begin{enumerate}
		\item Deviations that consistently mimic the behavior of another type. That is, some type $\theta$ behaves exactly as another type $\theta'$, i.e., $\beta_0'(\theta) = \beta_0(\theta')$ for some $\theta' \neq \theta$ and $\beta_1'(\theta,b_0,s) = \theta'$ for all $b_0 \in \supp \beta_0(\theta')$ and all $s$. The new constraint \eqref{E:ICB} ensures that such deviations are not profitable.
		\item Inconsistent deviations, i.e., mimic some other type $\theta'$ at time $0$  and report another type $\theta'' \neq \theta'$ at time $1$. Such deviations are off-path. Therefore we can replace the original mechanism $\gM$ with another mechanism $\gM'$ such that any inconsistent deviations are punished harshly enough so that they are never optimal and the equilibrium payoffs are unchanged. Thus we can assume that incentive constraints for such deviations are not binding.
		\item Finally, deviations to report $b_0$ that no type would choose in the original equilibrium. Again, such deviations are off-path and thus can be easily avoided, by replacing the original mechanism with a new one that excludes such unused messages from $\gB_0$.
	  \end{enumerate}	  
	\end{proof}		
This completes the proof of \cref{T:revelationprinciple}.
\end{proof}

\subsection{Proofs for \Cref{S:general}}
\subsubsection{Proof of \Cref{P:benchmark}} \label{SS:proof_benchmark}

\begin{proof}
Write the following problem 
\[
\max_{q_{0},t_{0}}\int_0^1 \int_0^1 \left[ t_{0}\left( \omega
,\theta \right) -\omega q_{0}\left( \omega ,\theta \right) \right] dG\left(
\omega \right) dF\left( \theta \right) 
\]

subject to%
\begin{eqnarray*}
\int_0^1 \left[ \theta q_{0}\left( \omega ,\theta \right) -t_{0}\left(
\omega ,\theta \right) \right] dG\left( \omega \right)  &\geq &0\text{ for
all }\theta \in \Theta  \\
\int_0^1 \left[ \theta q_{0}\left( \omega ,\theta \right) -t_{0}\left(
\omega ,\theta \right) \right] dG\left( \omega \right)  &\geq &\int_0^1 
\left[ \theta q_{0}\left( \omega ,\theta ^{^{\prime }}\right) -t_{0}\left(
\omega ,\theta ^{^{\prime }}\right) \right] dG\left( \omega \right) \text{
for all }\theta ,\theta ^{^{\prime }}\in \Theta 
\end{eqnarray*}%
Define%
\begin{eqnarray*}
Q_{0}\left( \theta \right)  &=&\int_ 0^1 q_{0}\left( \omega ,\theta
\right) dG\left( \omega \right)  \\
T_{0}\left( \theta \right)  &=&\int_0^1 t_{0}\left( \omega ,\theta
\right) dG\left( \omega \right)  \\
U_{0}\left( \theta \right)  &=&\theta Q_{0}\left( \theta \right)
-T_{0}\left( \theta \right) 
\end{eqnarray*}%
then the buyer's constraints become, respectively%
\begin{eqnarray*}
\theta Q_{0}\left( \theta \right) -T_{0}\left( \theta \right)  &\geq &0\text{
for all }\theta \in \Theta  \\
\theta Q_{0}\left( \theta \right) -T_{0}\left( \theta \right)  &\geq &\theta
Q_{0}\left( \theta ^{^{\prime }}\right) -T_{0}\left( \theta ^{^{\prime
}}\right) \text{ for all }\theta ,\theta ^{^{\prime }}\in \Theta 
\end{eqnarray*}%
These constraints can be replaced by%
\[
\begin{array}{c}
U_{0}\left( 0\right) =0\Leftrightarrow T_{0}\left( 0\right) =0 \\ 
Q_{0}\left( \theta \right) \text{ is weakly increasing in }\theta  \\ 
\frac{dU_{0}\left( \theta \right) }{d\theta }=Q_{0}\left( \theta \right)
\Leftrightarrow U_{0}\left( \theta \right) =\int_{0}^{\theta }Q_{0}\left(
y\right) dy\Leftrightarrow T_{0}\left( \theta \right) =\theta Q_{0}\left(
\theta \right) -\int_{0}^{\theta }Q_{0}\left( y\right) dy%
\end{array}%
\]%
Now%
\begin{align*}
&\int_0^1\int_0^1\left[ t_{0}\left( \omega ,\theta \right)
-\omega q_{0}\left( \omega ,\theta \right) \right] dG\left( \omega \right)
dF\left( \theta \right)  \\
&=\int_0^1 T_{0}\left( \theta \right) dF\left( \theta \right)
-\int_0^1 \omega q_{0}\left( \omega ,\theta \right) dG\left( \omega
\right) dF\left( \theta \right)  \\
&=\int_0^1 \left[ \theta Q_{0}\left( \theta \right) -\int_{0}^{\theta
}Q_{0}\left( y\right) dy\right] dF\left( \theta \right) -\int_0^1\int_0^1\omega q_{0}\left( \omega ,\theta \right) dG\left( \omega
\right) dF\left( \theta \right)  \\
&=\int_0^1 \int_0^1 (\theta - \omega) q_{0}\left( \theta ,\omega \right)
dG\left( \omega \right) dF\left( \theta \right) -\int_0^1\int_{0}^{\theta }\int_0^1q_{0}\left( y,\omega \right) dG\left(
\omega \right) dydF\left( \theta \right)  \\
&=\int_0^1 \int_0^1\left( \theta -\omega \right) q_{0}\left(
\theta ,\omega \right) dG\left( \omega \right) dF\left( \theta \right)
-\int_{0}^{1}\int_0^1 \int_{y}^{1}q_{0}\left( y,\omega \right) dF\left(
\theta \right) dG\left( \omega \right) dy \\
&=\int_0^1 \int_0^1\left( \theta -\omega \right) q_{0}\left(
\theta ,\omega \right) dG\left( \omega \right) dF\left( \theta \right)
-\int_0^1\int_0^1 q_{0}\left( \theta ,\omega \right) \left( \frac{%
1-F\left( \theta \right) }{f\left( \theta \right) }\right) dG\left( \omega
\right) dF\left( \theta \right)  \\
&=\int_0^1\int_0^1\left( \psi \left( \theta \right) -\omega
\right) q_{0}\left( \theta ,\omega \right) dG\left( \omega \right) dF\left(
\theta \right) ,
\end{align*}%
which is maximized by 
\[
q_{0}\left( \theta ,\omega \right) =\one\left[ \omega \leq \psi \left( \theta
\right) \right] 
\]%
and we can check that for $\theta \geq \psi^{-1}(0)$, $Q_{0}\left( \theta \right) =G\left( \psi\left(
\theta \right) \right)$ while for $\theta <  \psi^{-1}(0)$ we $Q_{0}\left( \theta \right) =0$ so that $Q_{0}\left( \theta \right)$ is weakly increasing.
 Now, for $\theta \geq \psi^{-1}(0)$
\[
T_{0}\left( \theta \right) =\theta G\left( \psi ^{+}\left( \theta \right)
\right) -\int_{0}^{\theta }G\left( \psi ^{+}\left( y\right) \right) dy
\]%
We can check that if $t_{0}\left( \theta ,\omega \right) =\psi ^{-1}\left(
\omega \right) 1\left[ \omega \leq \psi \left( \theta \right) \right] $ then
whenever $\theta \geq \psi ^{-1}\left( 0\right) $ where 
\begin{align*}
\int_{0}^{1}t_{0}\left( \theta ,\omega \right) dG\left( \omega \right)
&=\int_{0}^{1}\psi ^{-1}\left( \omega \right) \one\left[ \omega \leq \psi \left(
\theta \right) \right] dG\left( \omega \right)  \\
&=\int_{0}^{\psi \left( \theta \right) }\psi ^{-1}\left( \omega \right)
dG\left( \omega \right) =\theta G\left( \psi \left( \theta \right) \right)
-\int_{0}^{\psi \left( \theta \right) }G\left( \omega \right) d\psi
^{-1}\left( \omega \right)  \\
&=\theta G\left( \psi \left( \theta \right) \right) -\int_{0}^{\theta
}G\left( \psi \left( y\right) \right) dy
\end{align*}%
where the last integral obtains by defining $y=\psi ^{-1}\left( \omega
\right) .$ If $\theta <\psi ^{-1}\left( 0\right) $ then $t_{0}\left( \theta
,\omega \right) =T_{0}\left( \theta \right) =0$ for all $\omega $ and the proof is complete.
\end{proof}

\subsubsection{Proof of \Cref{C:properies_v}} \label{A:properies_v}

\begin{proof}
That $\lambda(\theta)$ is differentiable for $\theta > \theta^*$ follows immediately from inspection. Note that $\lambda(\theta)=0$ for all $\theta \leq \theta^*$ so the function is continuous and has a well defined right derivative at $\theta^*$. For the first property we have 
\begin{eqnarray*}
\int_{0}^{1}\lambda\left( \theta \right) dF\left( \theta \right) 
&=&\int_{0}^{1}\left( \theta -\psi ^{+}\left( \theta \right) \right) G\left(
\psi ^{+}\left( \theta \right) \right) dF\left( \theta \right)
-\int_{0}^{1}\int_{0}^{\theta }G\left( \psi ^{+}\left( y\right) \right)
dydF\left( \theta \right)  \\
&=&\int_{0}^{1}\left( \theta -\psi ^{+}\left( \theta \right) \right) G\left(
\psi ^{+}\left( \theta \right) \right) dF\left( \theta \right)
-\int_{0}^{1}\int_{y}^{1}G\left( \psi ^{+}\left( y\right) \right) dF\left(
\theta \right) dy \\
&=&\int_{0}^{1}\left( \theta -\psi ^{+}\left( \theta \right) \right) G\left(
\psi ^{+}\left( \theta \right) \right) dF\left( \theta \right)
-\int_{0}^{1}G\left( \psi ^{+}\left( \theta \right) \right) \left( \frac{%
1-F\left( \theta \right) }{f\left( \theta \right) }\right) dF\left( \theta
\right)  \\
&=&\int_{0}^{1}\left( \theta -\frac{1-F\left( \theta \right) }{f\left(
\theta \right) }-\psi ^{+}\left( \theta \right) \right) G\left( \psi
^{+}\left( \theta \right) \right) dF\left( \theta \right) \\
&=&\int_{\theta^*}^{1}\left( \theta -\frac{1-F\left( \theta \right) }{f\left(
\theta \right) }-\psi\left( \theta \right) \right) G\left( \psi
\left( \theta \right) \right) dF\left( \theta \right)=0
\end{eqnarray*}%
For the second property, we slightly abuse notation by writing $\lambda_{\theta}^{'}(\theta^*)$ for the right-derivative at $\theta^*$ and write:
\begin{eqnarray*}
\lambda_{\theta}^{'}(\theta^*) &=&G\left( \psi \left( \theta ^{\ast }\right) \right) \left(
1-\psi _{\theta }^{^{\prime }}\left( \theta ^{\ast }\right) \right) +\left(
\theta ^{\ast }-\psi \left( \theta ^{\ast }\right) \right) g\left( \psi
\left( \theta ^{\ast }\right) \right) \psi _{\theta }^{^{\prime }}\left(
\theta ^{\ast }\right) -G\left( \psi \left( \theta ^{\ast }\right) \right) 
\\
&=&\theta ^{\ast }g\left( \psi \left( \theta ^{\ast }\right) \right) \psi
_{\theta }^{^{\prime }}\left( \theta ^{\ast }\right) =\theta ^{\ast }g\left(
0\right) \psi _{\theta }^{^{\prime }}\left( \theta ^{\ast }\right) >0
\end{eqnarray*}%
since $\psi $ is strictly increasing. Finally,%
\begin{equation*}
\lambda\left( 1\right) =\left( 1-1\right) G\left( 1\right) -\int_{0}^{1}G\left(
\psi ^{+}\left( y\right) \right) dy=
-\int_{\theta ^{\ast }}^{1}G\left( \psi \left( y\right) \right) dy<0
\end{equation*}%
\end{proof}

\subsection{Proofs for \Cref{S:price}}

\subsubsection{Proof of \Cref{L:FREAisEAO}} \label{A:FREAisEAO}

\begin{proof}
 It is obvious that \eqref{E:optimalproblem} is a special case of \eqref{E:exanteproblem} with a fully revealing garbling device $m(\theta)=\theta$, therefore the solution of \eqref{E:optimalproblem} is within the constraint set of \eqref{E:exanteproblem} and the optimum must be weakly worse. For the other direction, note that when we denote $\oq_0(\theta,\omega)=q_0(m(\theta),\omega)$ and $\ot_0(\theta,\omega)=t_0(m(\theta),\omega)$, then we can rewrite \eqref{E:exanteproblem} as
\begin{equation} 
 \max_{(\oq_0,\ot_0)} \int_0^1 \int_0^1
 [\ot_0(\theta,\omega)-\omega \oq_0(\theta,\omega)]
 dG(\omega) dF(\theta)
\end{equation}
\begin{align*}
\int_0^1  [\theta \oq_0(\theta,\omega)-\ot_0(\theta,\omega)] dG(\omega) &\geq 0, \forall \theta
\\
\int_0^1  [\theta \oq_0(\theta,\omega)-\ot_0(\theta,\omega)] dG(\omega) 
&\geq 
\int_0^1  [\theta \oq_0(\theta',\omega)-\ot_0(\theta',\omega)] dG(\omega)
 , \forall \theta, \theta'
 \\
 \ot_0(\theta,\omega)- \omega \oq_0(\theta,\omega)
&\geq
 \ot_0(\theta,\omega')- \omega \oq_0(\theta,\omega')
 ,
\forall \theta, \omega, \omega'
\end{align*}
 subject to the constraint that $\oq_0$ and $\ot_0$ must be such that there exists a garbling device $m$ and functions $q_0$ and $t_0$ such that $\oq_0(\theta,\omega)=q_0(m(\theta),\omega)$ and $\ot_0(\theta,\omega)=t_0(m(\theta),\omega)$ for all $\theta,\omega$.
 Clearly, if we ignore this last constraint, we relax the problem. But this gives us \eqref{E:optimalproblem}.
\end{proof}

\subsubsection{Proof of \Cref{P:auxiliary_EAO}} \label{A:auxiliary_EAO}

\begin{proof}
Suppose we have a mechanism in $\gM^{EAP}$ that is fully-revealing and solves \eqref{E:optimalproblem}. We can thus focus on choosing $q_0(\theta,\omega)$ and $t_0(\theta,\omega)$.
By standard arguments, $q_0$ must be weakly decreasing in $\omega$. As $q_0$ is deterministic, for each $\theta$, there exists a marginal cost type $\uomega(\theta)$ such that $q_0(\theta,\omega)=\one[\omega \leq \uomega(\theta)]$. 
Let $\thetaone = \inf\{\theta: \uomega(\theta)=1\}$. By monotonicity, $\uomega(\theta)=1$ for all $\theta > \thetaone$ and $\uomega(\theta)<1$ for all $\theta<\thetaone$.
As $\theta=\thetaone$ is a zero-probability event, we can without loss of generality assume that $\uomega(\thetaone)=1$. 

From \eqref{E:ICS} we get that 
\[
  t_0(\theta,\omega) = \omega q_0(\theta,\omega) + \int_{\omega}^1 q_0(\theta,y) dy + t_0(\theta,1)-q_0(\theta,1)
\]
There are two cases. First, for $\theta < \thetaone$, $\uomega(\theta)<1$ and so $q_0(\theta,1)=0$. Then by the price mechanism assumption, $t_0(\theta,1)=0$. Hence:
\[
  t_0(\theta,\omega) 
  = \uomega(\theta) \one[\omega \leq \uomega(\theta)].
\]
This implies that for all $\theta < \thetaone$, $Q_0^S(\theta) = \int_0^1 q_0(\theta,\omega) dG(\omega) = G(\uomega(\theta))$ and thus
$T_0^{S}(\theta) = \int_0^1 t_0(\theta,\omega) dG(\omega) = \uomega(\theta) G(\uomega(\theta))$. On the other hand, from \eqref{E:IRB} and \eqref{E:ICB}, by standard arguments, we get that $Q_0^B(\theta)=G(\uomega(\theta))$ must be weakly increasing, so that $\uomega(\theta)$ must be weakly increasing, and we must have
\[
T_0^B(\theta) 
= \theta Q_0^B(\theta) - \int_0^{\theta} Q_0^B(x) dx
= \theta G(\uomega(\theta)) - \int_0^{\theta} G(\uomega(x)) dx. 
\]
Then $T_0^S(\theta)=T_0^B(\theta)$ gives us a condition
\[
[\theta-\uomega(\theta)] G(\uomega(\theta))
=
\int_0^{\theta} G(\uomega(x)) dx.
\]
This equality immediately implies that $\uomega(\theta) \leq \theta$ for all such $\theta$. 

We denote $p_0(\theta)=\uomega(\theta)$ for all $\theta < \thetaone$. Note that as types $\theta \geq \thetaone$ choose the guaranteed delivery option in equilibrium, we can set $p_0(\theta)$ for $\theta \geq \thetaone$ to any value, with a restriction that each type $\theta \geq \thetaone$ prefers to choose the guaranteed delivery option to any at-will offer  $p_0(\theta)$. One convenient way to guarantee this is to chose $p_0$ so that it satisfies the previous equation for all $\theta \in \Theta$. This is \eqref{E:auxpzerocond}.

Next, for all $\theta \geq \thetaone$, by construction $\uomega(\theta)=1$ so that $q_0(\theta,1)=q_0(\theta,\omega)=1$ for all $\omega$ and thus
\[
  t_0(\theta,\omega) = \omega + \int_{\omega}^1 dy + t_0(\theta,1)-1 = t_0(\theta,1).
\]
For all such $\theta$, $Q_0^B(\theta) = \int_0^1 q_0(\theta,\omega) dG(\omega) = 1$ and thus using \eqref{E:ICB} we get
\[
  T_0^B(\theta)
  = t_0(\theta,1)
  = \theta - \int_0^{\thetaone} Q_0^B(x) dx - \int_{\thetaone}^{\theta} 1 dx 
  = \thetaone - \int_0^{\thetaone} Q_0^B(x) dx = T_0^B(\thetaone).
\]
We denote $P_0 = T_0^B(\thetaone)$. Note, as there is a discontinous jump in the allocation rule at $\thetaone$, we need to determine this value directly from the \eqref{E:ICB} constraints. As the buyer type $\thetaone$ must be indifferent between guaranteed delivery at price $P$ and at-will offer at price $p_0(\thetaone)$, we get
\[
  \thetaone - P_0 = [\thetaone-p_0(\thetaone)] G(p_0(\thetaone))
  \iff 
  P_0 = p_0(\thetaone)  G(p_0(\thetaone)) + \thetaone [1-G(p_0(\thetaone))].
\]
This is \eqref{E:auxPcond}.

Finally, we need to show that $\thetaone < 1$. Suppose not, i.e., $\thetaone=1$. Then the profit from the optimal mechanism is 
\[
  V = \int_0^1 \int_0^{p_0(\theta)} (p_0(\theta)-\omega) dG(\omega) dF(\theta).
\]

Let us construct an alternative mechanism, with the same price function $p_0(\theta)$, but such that a guaranteed delivery offer $\hP$ is chosen by all types $1-\varepsilon$ for some small $\varepsilon>0$. By the analysis above, 
\[
  \hP = p_0(1-\varepsilon)  G(p_0(1-\varepsilon)) + (1-\varepsilon) [1-G(p_0(1-\varepsilon))].
\] 
Therefore the new profit can be computed as 
\begin{align*}
  \hV
  &= \int_0^{1-\varepsilon} \int_0^{p_0(\theta)} (p_0(\theta)-\omega) dG(\omega) dF(\theta) 
  + \int_{1-\varepsilon}^1 \int_0^1 (\hP-\omega) dG(\omega) dF(\theta)
\end{align*}
The difference between the two profits is
\begin{align*}
\hV-V
&= 
[1-F(1-\varepsilon)] \hP
-
\int_{1-\varepsilon}^1 p_0(\theta) G(p_0(\theta)) dF(\theta)
-
\int_{1-\varepsilon}^1 \int_{p_0(\theta)}^1 \omega dG(\omega) dF(\theta).
\end{align*}
As $p_0$ is weakly increasing, $p_0(\theta) G(p_0(\theta)) \leq p_0(1) G(p_0(1))$ for all $\theta \in [1-\varepsilon,1]$. Also by monotonicity of $p_0$, 
\[
  \int_{p_0(\theta)}^1 \omega dG(\omega) \leq \int_{p_0(1-\varepsilon)}^1 \omega dG(\omega).
\]
Combining these facts, we get 
\begin{align*}
\frac{\hV-V}{\varepsilon}
&\geq 
\frac{1-F(1-\varepsilon)}{\varepsilon} \left( \hP - p_0(1) G(p_0(1)) \right)
-
\frac{1-F(1-\varepsilon)}{\varepsilon} \int_{p_0(1-\varepsilon)}^1 \omega dG(\omega).
\end{align*}
Note that by \eqref{E:auxpzerocond}, $p_0(1)<1$. Taking limits as $\varepsilon \to 0$, we get that $\frac{1-F(1-\varepsilon)}{\varepsilon} \to f(1) > 0$, that
\[
  \hP - p_0(1) G(p_0(1))
  \to p_0(1)  G(p_0(1)) + 1-G(p_0(1))
  - p_0(1) G(p_0(1))
  = \int_{p_0(1)}^1 1 dG(\omega) > 0,
\]
and
\[
  \int_{p_0(1-\varepsilon)}^1 \omega dG(\omega)
  \to 
  \int_{p_0(1)}^1 \omega dG(\omega) > 0.
\]
Combining these facts, 
\[
\lim_{\varepsilon \to 0}
\frac{\hV(\varepsilon)-V}{\varepsilon}
= f(1) 
  \int_{p_0(1)}^1 (1-\omega) dG(\omega) > 0.
\]
Therefore, for a sufficiently small $\varepsilon > 0$, $\hV(\varepsilon)>V$, which is a contradiction with the assumption that the original mechanism is optimal. This proves that, $\thetaone < 1$, i.e., the guaranteed delivery offer is chosen with positive probability. 
\end{proof}	

\subsubsection{Proof of \Cref{C:EAOvsB}} \label{A:EAOvsB}

\begin{proof}
As discussed above, the profit from any mechanism satisfying \eqref{E:IRB} and \eqref{E:ICB}, can be computed as 
\[
  V = \int_0^1 \int_0^1 [\psi(\theta)-\omega] q(\theta,\omega) dG(\omega) dF(\theta).
\]
Specifically, the profit from the benchmark mechanism is
\[
  V^B = \int_0^1 \int_0^{\psi(\theta)} (\psi(\theta)-\omega) dG(\omega) dF(\theta)
\]	
and the profit from the optimal ex-ante price mechanism is
\[
  V^{EAO}
  = \int_0^{\thetaone} \int_0^{p_0(\theta)} (\psi(\theta)-\omega) dG(\omega) dF(\theta)
  + \int_{\thetaone}^1 \int_0^1 (\psi(\theta)-\omega) dG(\omega) dF(\theta).
\]
Note that as for all $\theta$,
\[
\int_0^{\psi(\theta)} (\psi(\theta)-\omega) dG(\omega)
\geq 
\int_0^{p_0(\theta)} (\psi(\theta)-\omega) dG(\omega),
\]
the difference 
\[
  V^B 
  -
  V^{EAO}
  \geq 
  -\int_{\thetaone}^1 \int_{\psi(\theta)}^1 (\psi(\theta)-\omega) dG(\omega) dF(\theta) > 0,  
\]
as $\thetaone < 1$.
\end{proof}	
\subsubsection{Proof of \Cref{C:EAOvsEASI}} \label{A:EAOvsEASI}
\begin{proof}
We can write the expected profits for the seller with a guaranteed delivery price $P_0$ as
\begin{align*}
\left[1 - F(P_0)\right] \left(P_0 - \expect[\omega]\right)
= \left[1 - F(P_0)\right] \left(P_0 - 1 + \gG(1) \right)].
\end{align*}

Now, the profits for a mechanism where we still have a single price $P_0$ but with \emph{at-will} delivery are 
\begin{align*}
\int_{P_0}^1 \int_0^{P_0} (P_0 - \omega)dG(\omega)dF\theta)= \left[1 - F(P_0)\right] \gG(P_0).
\end{align*}
The difference between the latter and the former is therefore:
\begin{align*}
  \left[1 - F(P_0)\right] \left[ \gG(P_0)-P_0+1-\gG(1)\right]=\left[1 - F(P_0)\right] \int_{P_0}^1 (1-G(x))dx>0.
\end{align*}

So, for any $P_0$ that maximizes the seller's profits with guaranteed delivery, the seller can do better with a mechanism where she uses the same price but with at-will delivery. But this is bettered, as established by \Cref{P:auxiliary_EAO}, by a menu of offers and the result follows. 
\end{proof}
\subsubsection{Proof of \Cref{P:DvsEPO}} \label{A:DvsEPO}

\begin{proof}
	Clearly, buyers with highest types have the strongest incentive to accept the first offer, so there must be a marginal type $\otheta$ who is indifferent between accepting and rejecting the first offer.

	In this mechanism, if the buyer did not buy at time 0, the seller at time 1 faces a buyer $\theta \in [0,\otheta]$ with a cumulative distribution function $F(\theta|\theta \leq \otheta) =\frac{F(\theta)}{F(\otheta)}$. Following the usual steps we can write the seller's expected profits at time 1, as a function of the probability of trade, and given the buyer's type is in $[0,\otheta]$ as
	\[ 
	  V_1^{D}(\otheta) =
	  \int_0^{\otheta} 
		  \int_0^{\otheta} 
			 \left [ q_1(\theta,\omega) - \int_0^\theta q_1(y,\omega)dy - \omega q_1(\theta,\omega) \right ] \frac{f(\theta)}{F(\otheta)} d\theta dG(\omega),
	\]
	so that the expected profits for the seller at time 0 in the D mechanism, given $\otheta$, are
	
	\[ 
	  V^{D}(\otheta, \delta) = (1 - F(\otheta))
	  \int_0^1 (p_0 - \omega ) dG(\omega) + \delta F(\otheta)V_1^{D}(\otheta), 
	\]
	where $p_0$ solves, for a fixed $\otheta$, the equation
	\begin{equation} \tag{CO} \label{E:Coasian}
	  \otheta - p_0 = \delta \int_0^{\otheta} 
		  \int_0^{\otheta} 
			  q_1(y,\omega) dy dG(\omega), 
	\end{equation}
	in which the right-hand side represents the expected utility for the type $\otheta$ buyer at time 1 if she rejects price $p_0$. \Cref{E:Coasian} captures the Coasian constraint that types $\theta \in [\otheta,1]$ prefer to accept price $p_0$ to the expected utility they could get if they waited for the mechanism offered at time 1. Now, since at time 1, the seller learns $\omega$, we can apply the standard steps to get
	$
	  q_1(\theta,\omega)
	  = \mathbf{1} \left[\omega < \psi(\theta,\otheta)\right],
	$
	where $\psi(\theta,\otheta) = \theta - \frac{F(\otheta)- F(\theta)}{f(\theta)}$, 
	so that
	\[ 
	  p_0 = \otheta - \delta \int_0^{\otheta} 
		  \int_0^{\max(0,\psi(\theta,\otheta))} 
			 dG(\omega)dy = \otheta - \delta \int_{\theta^{**}(\otheta)}^{\otheta} 
		  G(\psi(\theta,\otheta)) d\theta , 
	\]
	where $\theta^{**}(\otheta)$ is the value of $\theta$ for which $\psi(\theta,\otheta) = 0$. If $\psi(\theta,1)$ is strictly increasing in $\theta$ then so is $\psi(\theta,\otheta)$, which implies that $\theta^{**}(\otheta)$ is well-defined, and $\theta^{**}(1) = \theta^*$.
	Also,
	\begin{align*}
	  V_1^{D}(\otheta)F(\otheta) 
		&= \int_0^{\otheta} 
		  \int_0^{\otheta} 
			 \left [ q_1(\theta,\omega) - \int_0^\theta q_1(y,\omega)dy - \omega q_1(\theta,\omega) \right ] f(\theta) d\theta dG(\omega) \\
		&= \int_{\theta^{**}(\otheta)}^{\otheta} 
		 \left (\theta f(\theta) - F(\otheta) + F(\theta) \right) G(\psi(\theta,\otheta)) d\theta 
		 \\
		 &- \int_{\theta^{**}(\otheta)}^{\otheta} 
		 f(\theta) \left[ \psi(\theta,\otheta))G(\psi(\theta,\otheta)) - \gG(\psi(\theta,\otheta))\right] d\theta \\
		 &= \int_{\theta^{**}(\otheta)}^{\otheta} 
		 \gG(\psi(\theta,\otheta)) f(\theta) d\theta,
	\end{align*}
	Hence, 
	\[
	  V^{D}(\otheta, \delta) = (1 - F(\otheta))
	  \left [\otheta - \delta \int_{\theta^{**}(\otheta)}^{\otheta} 
		 G(\psi(\theta,\otheta)) d\theta - \expectG[\omega] \right] + \delta \int_{\theta^{**}(\otheta)}^{\otheta} 
		 \gG(\psi(\theta,\otheta)) f(\theta) d\theta.
	\]
	Recalling that $\expectG[\omega] = \int_0^1 \omega dG(\omega) = 1 - \gG(1)$, we now consider 
	\begin{align*}
	  \frac{\partial V^{D}(\otheta,\delta) }{\partial \otheta} 
	  &= -f(\otheta)
	  \left [\otheta - \delta \int_{\theta^{**}(\otheta)}^{\otheta} 
		 G(\psi(\theta,\otheta)) d\theta - 1 +\gG(1) \right ] \\
		 &+ (1-F(\otheta)) \left [1- \delta\frac{d \int_{\theta^{**}(\otheta)}^{\otheta}G(\psi(\theta,\otheta))d\theta}{d\otheta}\right] 
		 \\
		 &+ \delta f(\otheta)) \left [\gG(\psi(\otheta,\otheta)) -  \int_{\theta^{**}(\otheta)}^{\otheta} 
		 G(\psi(\theta,\otheta)) d\theta \right], 
	\end{align*} 
	and if we evaluate this at $\otheta = 1$ we get 
	\begin{align*}
	  \frac{\partial V^{D}(1,\delta) }{\partial \otheta} &= -f(1)
	  \left [1 - \delta \int_{\theta^*}^1 
		 G(\psi(\theta,1)) d\theta - 1 +\gG(1) \right ] 
		 \\
		 &+(1-F(1)) \left[ 1 - \delta \xi'(1) \right]
		 + \delta f(1) \left [\gG(1) -  \int_{\theta^*}^1 
		 G(\psi(\theta,1)) d\theta  \right] \\
		 &= f(1) \left[\delta \int_{\theta^*}^1 
		 G(\psi(\theta,1)) d\theta -\gG(1) + \delta \gG(1) - \delta \int_{\theta^*}^1 
		 G(\psi(\theta,1)) d\theta  \right] \\
		 &= f(1) \gG(1) (\delta -1),  
	\end{align*} 
	where
	$\xi(\otheta) = \int_{\theta^{**}(\otheta)}^{\otheta} G(\psi(\theta,\otheta)) d\theta$, so that
	$\xi'(1)
	=
	1
	-
	f(1) \int_{\theta^*}^{1} \frac{g(\psi(\theta,1))}{f(\theta)} d\theta
	\leq 1 < \infty$ and therefore $(1-F(1))\xi'(1)=0$.
	So, for any $\delta < 1$, $V^{D}(\otheta,\delta)$ is decreasing at $\otheta =1$ and this implies that it is optimized for $\otheta < 1$. Since $V^{D}(1,\delta) = V^{EPO}$ then this proves the result. 
	\end{proof}

\subsection{Proofs for \Cref{S: extensions}} \label{S:proofs_extensions}
\subsubsection{Proof of \Cref{P: necessary_learning_benchmark}} \label{SS: proof_necessary_learning_benchmark}
\begin{proof}
Define the set 
\begin{gather*}
    \widehat{\Theta}(c)=\{\theta: \exists \: 
 \omega \: \text{such that} \:
 q_0(\theta,\omega)>0\},
\end{gather*}
then the problem faced by the seller is a simple modification of the benchmark problem where the constraints for the buyer are the same but the objective function is now:
\begin{gather*}
\int_0^1 \int_0^1 [ t_0(\theta,\omega)-\omega q_0(\theta,\omega)] dF(\theta) dG(\omega) - \int_{\widehat{\Theta}(c)} c dF(\theta).
\end{gather*}
Since the buyer's constraints are unchanged, using the same procedure as in \cref{SS:proof_benchmark} we can rewrite the objective as
\begin{gather*}
\int_0^1 \int_0^1 [ \psi(\theta)-\omega ]q_0(\theta,\omega) dF(\theta) dG(\omega) - \int_{\widehat{\Theta}(c)} c dF(\theta).
\end{gather*}

As the seller observes $\theta$ before choosing whether the learn her cost, we can maximize this for any given $\theta$. That is, for any type $\theta$, the seller will have to decide whether to make trade possible for some $\omega$. This will require spending the learning cost $c$. The other option is not to spend the learning cost and not trading. Hence, the optimal $q_0$ maximizes:
\begin{equation*}
\max \left(\int_0^1 [ \psi(\theta)-\omega - c]q_0(\theta,\omega) dG(\omega),0 \right) \; \forall \theta \in \Theta.
\end{equation*}
This immediately gives us:
\begin{align*}
    q_0(\theta,\omega)
    &= \one[\omega \leq\ \psi(\theta)]
    \cdot \one\left[\int_0^{\psi(\theta)}(\psi(\theta) - \omega -c) dG(\omega) \geq 0\right] \\
    &=\one[\omega \leq\ \psi(\theta)]
    \cdot \one\left[\ \gG(\psi(\theta)) \geq c\right] = \one[\max(\omega,\gG^{-1}(c)) \leq \psi(\theta)].
\end{align*}
It is easy to see that this means that $q_0=0$ whenever $c > \gG(1)$. The rest follows as in the benchmark case noting that $\widehat{\Theta}(c)=\{\theta: \theta \geq \psi^{-1}(\gG^{-1}(c))\}$. 
\end{proof}

\subsubsection{Proof of \Cref{P: optional_learning_benchmark}} \label{SS: proof_optional_learning_benchmark}
\begin{proof}
The proof follows along the lines of the proof of \cref{P: necessary_learning_benchmark}. So define $\overline{\Theta}(c)$ as the set of buyer types for which the seller will be willing to invest in learning her type. Contrary to the proof of \cref{P: necessary_learning_benchmark}, however, we will not define it ex-ante as we did with $\widehat{\Theta}$. It is immediate, however, $\overline{\Theta}(c) 
\subseteq \widehat{\Theta}(c)$ for any $c$, as the seller no longer needs to learn her cost in order to trade. Given this, and given that again we are maximizing for any given $\theta$, we have that whenever $\theta \in \overline{\Theta}(c)$, the seller's expected utility taking into account all the constraints for the buyer is:  
\begin{gather*}
\int_0^1 [q_0(\theta,\omega)(\psi(\theta)-\omega) - c] dG(\omega)=\int_0^1 q_0(\theta,\omega)(\psi(\theta)-\omega) dG(\omega) -c,
\end{gather*}
which, as we've seen in the proof of proposition \cref{P: necessary_learning_benchmark} is optimized by $q_0(\theta,\omega)=\one[\max(\omega,\gG^{-1}(c)) \leq \psi(\theta)]$. 
If $\theta \notin \overline{\Theta}(c)$ we have:
\begin{gather*}
\int_0^1 q_0(\theta)(\psi(\theta)-\omega) dG(\omega),
\end{gather*}
where, with a slight abuse of notation, we have emphasized that in this latter case, $q_0$ cannot be a function of $\omega$ as, by definition, this is not learned. This is clearly maximized by $q_0(\theta)=q_0(\theta,\omega)=\one[\mathbb{E}(\omega) \leq \psi(\theta)]$. 

For any $\theta$ such that the mechanism gives no probability of trade, it is clearly optimal not to learn $\omega$. Therefore, we now assume that $\theta \geq \psi^{-1}(0)$ as this is the largest possible set of buyer types with whom the seller may wish to trade.
 For any such $\theta$, define the function:
\begin{align*}
\chi(\theta)
&= \int_0^{\psi(\theta)} (\psi(\theta)-\omega) dG(\omega) - \int_0^1 (\psi(\theta)-\omega) dG(\omega)=-\int_{\psi(\theta)}^1 (\psi(\theta)-\omega) dG(\omega) \\
&=\mathbb{E}(\omega)+\gG(\psi(\theta))-\psi(\theta).
\end{align*}
The $\chi(\theta)$ function tells us, for any $\theta$ for which the mechanism gives a positive probability of trade, the gain for the seller from learning her cost. It is easy to check that for any $\theta \in [\psi^{-1}(0),1)$ this is a positive function because $\mathbb{E}(\omega)=1-\gG(1)$ so that
\begin{equation*}
    \chi(\theta)=\int_{\psi(\theta)}^1 [1-G(x)]dx>0.
\end{equation*}
Further, $\chi(\psi^{-1}(0))=\mathbb{E}(\omega)$, $\chi(1)=0$, and $\chi(\theta)$ is strictly decreasing. 
The seller will wish to learn her type for any $\theta$ such that $\chi(\theta) \geq c$ \emph{and} $\gG(\psi(\theta)) \geq c$. The latter condition tells us that for this $\theta$, the seller prefers to trade and learn her type over not trading, whereas the former guarantees that she also prefers to trade learning her type (and paying the cost of learning) over trading without learning her type.
The function $\gG(\psi(\theta))$ is strictly positive in the interval $\theta \in (\psi^{-1}(0),1]$, and is strictly increasing with $\gG(\psi(\psi^{-1}(0)))=0$ and $\gG(\psi(1))=\gG(1)$.
Thus, the two functions must cross exactly once at some point $\widehat{\theta}$. But if $\chi(\widehat{\theta})=\gG(\psi(\widehat{\theta}))$, then it is immediate to see that is must also be that $\psi(\widehat{\theta)}=\mathbb{E}(\omega) \iff \widehat{\theta}=\psi^{-1}(\mathbb{E}(\omega))$. 

Now, fix $c$. Given the above, the set $\overline{\Theta}(c)$ is non-empty whenever $c \leq \gG(\mathbb{E}(\omega))$. If so,
\begin{gather*}
\overline{\Theta}(c)=\{\theta: \chi(\theta) \geq c \; \text{and} \; \gG(\psi(\theta)) \geq c \} = [\psi^{-1}(\gG^{-1}(c)),\chi^{-1}(c)],
\end{gather*}
with $\widehat{\theta} \in [\psi^{-1}(\gG^{-1}(c)),\chi^{-1}(c)]$. Clearly, $\psi^{-1}(\gG^{-1}(c))$ is strictly increasing in $c$ with $\psi^{-1}(\gG^{-1}(0))=\psi^{-1}(0)$ and $\psi^{-1}(\gG^{-1}((\mathbb{E}(\omega))))=\widehat{\theta}$ while $\chi^{-1}(c)$ is strictly decreasing with $\chi^{-1}(0)=1$ and $\chi^{-1}(\gG(\mathbb{E}(\omega))=\widehat{\theta}$   

If $c > \gG(\mathbb{E}(\omega))$, then for all $\theta \geq \psi^{-1}(0)$, at least one of the two conditions for the seller wanting to learn her type is not satisfied. The seller then, knows ex-ante that she will never want to learn her type and chooses a mechanism that does not depend on it, but only on the buyer's type. The optimal mechanism is then to trade whenever the buyer's virtual valuation is no smaller than the seller's \emph{expected} type, as shown above. The corresponding transfers follow immediately as in \cref{P:benchmark} with $\mathbb{E}(\omega)$ instead of $\omega$. 

If $c \leq \gG(\mathbb{E}(\omega))$, instead, the interval is well-defined and so the seller will choose the allocation rule $q_0(\theta,\omega)=\one[\max(\omega,\gG^{-1}(c)) \leq \psi(\theta)]$ whenever $\theta \in \overline{\Theta}(c)$ which, by construction of $\overline{\Theta}(c)$, becomes $\one[\omega \leq \psi(\theta)]$. If $\theta > \chi^{-1}(c)$, because this implies $\theta > \widehat{\theta}=\psi^{-1}(\mathbb{E}(\omega))$, the seller will always trade the good. Noting that the monotonicity conditions on either  $q_0(\theta)$ (when there is no learning) or $Q_0^B(\theta), \; q_0(\theta,\omega)$ (when there is learning) apply and this proves the optimal allocation. 

As for transfers, given that our constraints only put restrictions on $T_0^B(\theta)$, it is without loss to have $t_0(\theta,\omega)=T_0^B(\theta)$ whenever a buyer type has probability 0 or 1 of trade and to set \begin{equation*}
t_0(\theta,\omega)=\frac{T_0^B(\theta)}{G(\psi(\theta))} \one[\omega \leq \psi(\theta)],
\end{equation*}
whenever trade also depends on the realization of $\omega$ (that is, whenever $\theta \in \overline{\Theta}(c)$). Given the buyer's constraints we then have that $T_0^B(\theta)=\theta Q_0^B(\theta)-\int_0^{\theta}Q_0^B(y)dy$. It is immediate that if $\theta < \psi^{-1}(\gG^{-1}(c))$ then $t_0(\theta,\omega)=T_0^B(\theta)=0$, whereas if $\theta > \chi^{-1}(c)$ then $t_0(\theta,\omega)=T_0^B(\theta)=\chi^{-1}(c)-\int_{\psi^{-1}(\gG^{-1}(c))}^{\chi^{-1}(c)}G(\psi(y))dy$. Finally, if $\theta \in \overline{\Theta}(c)$, then $T_0^B(\theta)=\theta - \int_{\psi^{-1}(\gG^{-1}(c))}^{\theta}G(\psi(y))dy$ which immediately implies the transfers $t_0(\theta,\omega)=\frac{T_0^B(\theta)}{G(\psi(\theta))}\one[\omega \leq \psi(\theta)]$ given in the proposition.

The formulas for $V^{BOC}$ are immediate. As for the comparison with $V^{BNC}$, note that if $c>\gG(1)$ then the former is positive (the seller offers trade with guaranteed delivery at a fixed price $\psi^{-1}(\mathbb{E}(\omega)$) while the latter is zero (the seller doesn't trade). If $\gG(1) \geq c >\gG(\mathbb{E}(\omega))$ then we have that $\psi^{-1}(\gG^{-1}(c))$ is still well-defined and, given the above:  
\begin{align*}
    V^{BOC}&=\int_{\psi^{-1}(\mathbb{E}(\omega))}^1 \int_0^1 
 \left(
    \psi(\theta) - \omega  
  \right)  dG(\omega) dF(\theta)
\\  &>\int_{\psi^{-1}(\gG^{-1}(c))}^1 \int_0^{\psi(\theta)}  
 \left(
    \psi(\theta) - \omega  -c
  \right)  dG(\omega) dF(\theta)  \\
 &> \int_{\psi^{-1}(\gG^{-1}(c))}^1 -c+ \left(\int_0^{\psi(\theta)}  
 \left(
    \psi(\theta) - \omega  -c
  \right)  dG(\omega) \right) dF(\theta)=V^{BNC},
  \end{align*} 
where the first inequality comes from the fact that since $c >\gG(\mathbb{E}(\omega))$, then when $\gG(\psi(\theta)) \geq c$, necessarily $\chi(\theta) < c$.
Finally, if $c \leq \gG(\mathbb{E}(\omega))$ then
\begin{align*}
    V^{BOC}
    &>\int_{\psi^{-1}(\gG^{-1}(c))}^1  \int_0^{\psi(\theta)} (\psi(\theta) - \omega -c)dG(\omega) dF(\theta) \\ 
    &>\int_{\psi^{-1}(\gG^{-1}(c))}^1 \left( -c+ \int_0^{\psi(\theta)} (\psi(\theta) - \omega)dG(\omega) \right)dF(\theta)=V^{BNC}.
\end{align*}
Now, the first inequality comes from simple inspection of $V^{BOC}$.    
\end{proof}

\subsubsection{Proof of \Cref{P: efficiency}} \label{SS: proof_efficiency}
\begin{proof}
We only need to check that the ex-post optimal allocation and the transfers   $t_0(\theta,\omega)=\theta \one[\omega \leq \theta]-\gG(\theta)$ satisfy all the constraints. The usual monotonicity conditions on $Q_0^B(\theta)$ and $q_0(\theta,\omega)$ are clearly satisfied. We then have
\begin{align*}
T_0^S(\theta)
&=\int_0^1 t_0(\theta,\omega)dG(\omega)=v(\theta,1)+\int_0^1 \left( \int_{\omega}^1 \one[x \leq \theta]dx+\omega \one[\omega \leq \theta] \right)dG(\omega) \\
&= \lambda^E(\theta)+\int_0^1 \one[\omega \leq \theta] G(\omega)d\omega + \int_0^1 \omega \one[\omega \leq \theta] dG(\omega) \\
&= \lambda^E(\theta)+\gG(\theta)+\theta G(\theta)-\gG(\theta) = \lambda^E(\theta)+\theta G(\theta),
\end{align*}
whereas
\begin{align*}
T_0^B(\theta)
&=\theta Q_0^B(\theta)-\int_0^{\theta}Q_0^B(y)dy=\theta \int_0^1 \one[\omega \leq \theta]dG(\omega) - \int_0^{\theta}\int_0^1 \one[\omega \leq y]dG(\omega)dy \\
&=\theta G(\theta) - \int_0^{\theta}G(y)dy=\theta G(\theta) - \gG(\theta),
\end{align*}
and, of course, equating them gives us $\lambda^E(\theta)=-\gG(\theta)$. To check that this is individually rational for the seller we can show that
\begin{align*}
    \int_0^1 v(\theta,\omega)dG(\omega)
    &=\int_0^1 [t_0(\theta,\omega)-\omega q_0(\theta,\omega)]dG(\omega)=\int_0^{\theta}(\theta - \omega)dG(\omega) - \gG(\theta) \\
    &=\theta G(\theta) - [\theta G(\theta) -\gG(\theta)] - \gG(\theta)=0,
\end{align*}
which means that the ``reverse" individual rationality condition for the seller is satisfied with equality for all buyer types. This, of course, also implies the seller's ex-ante individual rationality.
\end{proof}

\end{document}